\documentclass[preprint,showpacs,preprintnumbers,amsmath,amssymb]{revtex4}

\usepackage{graphicx}% Include figure files
\usepackage{dcolumn}% Align table columns on decimal point
\usepackage{bm}% bold math

\begin{document}

\preprint{APS/123-QED}
\def\a{\acute{a}}
\def\b{\acute{b}}
\def\c{\acute{c}}
\def\d{\acute{d}}

\def\x{\acute{c}}

\def\beq{\begin{equation}}
\def\eeq{\end{equation}}
\def\beqn{\begin{eqnarray}}
\def\eeqn{\end{eqnarray}}
\def\rmuu{\gamma^{\mu}}
\def\rmud{\gamma_{\mu}}
\def\PL{{1-\gamma_5\over 2}}
\def\PR{{1+\gamma_5\over 2}}
\def\sinW2{\sin^2\theta_W}
\def\AEM{\alpha_{EM}}
\def\mul{M_{\tilde{u} L}^2}
\def\mur{M_{\tilde{u} R}^2}
\def\mdl{M_{\tilde{d} L}^2}
\def\mdr{M_{\tilde{d} R}^2}
\def\mz2{M_{z}^2}
\def\c2b{\cos 2\beta}
\def\au{A_u}
\def\ad{A_d}
\def\cob{\cot \beta}
\def\v#1{v_#1}
\def\tb{\tan\beta}
\def\epem{$e^+e^-$}
\def\KK{$K^0$-$\overline{K^0}$}
\def\wi{\omega_i}
\def\xj{\chi_j}
\def\Wmu{W_\mu}
\def\Wnu{W_\nu}
\def\m#1{{\tilde m}_#1}
\def\mH{m_H}
\def\mw#1{{\tilde m}_{\omega #1}}
\def\mx#1{{\tilde m}_{\chi^{0}_#1}}
\def\mc#1{{\tilde m}_{\chi^{+}_#1}}
\def\mwi{{\tilde m}_{\omega i}}
\def\mxi{{\tilde m}_{\chi^{0}_i}}
\def\mci{{\tilde m}_{\chi^{+}_i}}
\def\mz{M_z}
\def\sw{\sin\theta_W}
\def\cw{\cos\theta_W}
\def\cb{\cos\beta}
\def\sb{\sin\beta}
\def\rwi{r_{\omega i}}
\def\rxj{r_{\chi j}}
\def\rfp{r_f'}
\def\Kik{K_{ik}}
\def\Fq2{F_{2}(q^2)}
\def\f{\({\cal F}\)}
\def\d1{{\f(\tilde c;\tilde s;\tilde W)+ \f(\tilde c;\tilde \mu;\tilde W)}}
%%%%%%%%%%%%%%%%%%%%%%%%%%%%%%%%%%
\def\tw{\tan\theta_W}
\def\sec2w{sec^2\theta_W}
\newcommand{\lLambda}{\mbox{\Large$\Lambda$}}
\newcommand{\LLambda}{\mbox{\LARGE$\Lambda$}}
\newcommand{\blLambda}{\mbox{\Large\boldmath$\Lambda$}}
\newcommand{\bLambda}{\mbox{\boldmath$\Lambda$}}
\newcommand{\bXi}{\mbox{\boldmath$\Xi$}}
\newcommand{\blXi}{\mbox{\Large\boldmath$\Xi$}}
\newcommand{\blOmega}{\mbox{\Large\boldmath$\Omega$}}
\newcommand{\bDelta}{\mbox{\boldmath$\Delta$}}
\newcommand{\blDelta}{\mbox{\Large\boldmath$\Delta$}}
\newcommand{\blPsi}{\mbox{\Large\boldmath$\Psi$}}
\newcommand{\ltau}{\mbox{\Large$\tau$}}
\newcommand{\lmu}{\mbox{\Large$\mu$}}
\newcommand{\lnu}{\mbox{\Large$\nu$}}
\newcommand{\Nu}{\mbox{\LARGE$\nu$}}
\newcommand{\Tau}{\mbox{\LARGE$\tau$}}
\newcommand{\bNu}{\mbox{\boldmath$\nu$}}
\newcommand{\bTau}{\mbox{\boldmath$\tau$}}
\newcommand{\bnu}{\mbox{\boldmath$\nu$}}
\newcommand{\bmu}{\mbox{\boldmath$\mu$}}
\newcommand{\bgamma}{\mbox{\boldmath$\gamma$}}
\newcommand{\blpi}{\mbox{\Large\boldmath$\pi$}}
\newcommand{\blrho}{\mbox{\Large\boldmath$\rho$}}
\newcommand{\brho}{\mbox{\boldmath$\rho$}}
\newcommand{\bleta}{\mbox{\Large\boldmath$\eta$}}
\newcommand{\bpi}{\mbox{\boldmath$\pi$}}
\newcommand{\mTau}{\mbox{\normalsize$\tau$}}
\newcommand{\mNu}{\mbox{\normalsize$\nu$}}
%%%%%%%%%%%%%%%%%%%%%%%%%%%%%%%%%%
\title{Suppression of Higgsino mediated proton decay
by cancellations
 in GUTs and strings}% Force line breaks with \\

\author{Pran Nath }
\email{nath@lepton.neu.edu}
\author{Raza M. Syed}
\email{r.syed@neu.edu}
\affiliation{Department of Physics, Northeastern University,
Boston, MA 02115-5000, USA. }

\date{\today}% It is always \today, today,
             %  but any date may be explicitly specified

\begin{abstract}
A  mechanism for the enhancement for proton
 lifetime in supersymmetric/supergravity  (SUSY/SUGRA)   grand unified theories (GUTs)
 and in string theory models  is discussed
 where Higgsino mediated proton decay
  arising from color triplets (anti-triplets) with charges  $Q=-1/3(1/3)$
 and $Q=-4/3(4/3)$ is suppressed by an internal cancellation due to
 contributions from different sources. We exhibit  the mechanism
 for an $SU(5)$ model with  $45_H+\overline{45}_H$ Higgs
 multiplets in addition to the usual Higgs structure of the minimal model.
 This model contains both $Q=-1/3(1/3)$ and $Q=-4/3(4/3)$ Higgs color
 triplets (anti-triplets) and simple constraints allow for  a complete suppression
 of Higgsino mediated proton decay. Suppression of proton decay in an
  $SU(5)$ model with Planck scale contributions  is also considered.
 The suppression  mechanism is then exhibited  for an $SO(10)$ model with
a  unified Higgs structure involving $144_H+\overline{144}_H$
representations.
 The $SU(5)$ decomposition of $144_H+\overline{144}_H$ contains $5_H+\bar 5_H$
 and $45_H+\overline{45}_H$ and the cancellation mechanism arises among these
 contributions which mirrror the $SU(5)$ case.
The cancellation  mechanism appears to be more generally valid for
a larger  class of  unification models.
Specifically the cancellation mechanism may
play a role in string model constructions to suppress proton decay
from dimension five  operators.
The  mechanism allows for
the suppression of proton decay consistent with current data
allowing for the possibility that  proton decay may be visible in
the next round of nucleon stability experiment. \end{abstract}
\pacs{12.10.Dm; 13.30.- a; 12.10.- g; 11.25.Wx; 11.30.Pb; 12.60.- i}
\maketitle
\section{\label{sec1}INTRODUCTION}
Grand unification and strings are attractive schemes for the
unification of interactions. One consequence of grand unification
is that one has baryon and lepton number non-conservation which
can lead  to proton decay\cite{Pati:1974yy,Georgi:1974sy,georgi}
and a similar phenomenon occurs in string models (for a review see
\cite{Nath:2006ut}). In supersymmetric and supergravity  GUT
 theories\cite{Dimopoulos:1981zb,Chamseddine:1982jx}
  baryon and lepton number  violating dimension five operators are
the dominant contributions\cite{pdecay1,pdecay2}. These
contributions are now stringently constrained by experiment.
Thus the analysis of Ref.\cite{mp} indicates that the minimal
$SU(5)$ even in the decoupling limit is eliminated\cite{mp}
(we note in passing that the minimal $SU(5)$ model is eliminated in any
case  since it fails to reproduce the fermion masses)
and
further that supersymmetric grand unification in general  may also
be under siege\cite{dmr} due to the current experimental lower
limits on the proton lifetime\cite{Kobayashi:2005pe,Yao:2006px}.
 While there are ways to lift the siege (see, e.g., \cite{bps})
the proton lifetime limit is  certainly one of the  most important
constraints on grand unification and on string models, and
 is likely to become even more
stringent as the error corridor on the predictions decrease and
the lower limits from experiment improve. Several possible avenues
for suppressing  proton decay have already been discussed in the
literature from mild suppression using textures\cite{textures},
 and   CP phases\cite{cp} to stronger suppression\cite{suppress}, and suppression
 up to the current limit of experiment \cite{Dutta:2004zh}.
Here we add to this list the cancellation mechanism for the
suppression of proton decay from dimension five operators  which
is inspired by a similar mechanism used to suppress the EDM of the
electron and of the neutron in supersymmetric
theories\cite{Ibrahim:1998je}.

The outline of the rest of the paper is as follows: in
Sec.~\ref{sec2} we discuss the constraints necessary for the
suppression of baryon and lepton number violating dimension five
operators arising from Higgsino exchange. These constraints are
valid both for grand unified theories as well as for models
arising from strings. In Sec.~\ref{sec3} we discuss an $SU(5)$
grand unification model where we include a
 $45 +\overline{45}$ plet of Higgs in addition to the usual Higgs structure of the
 minimal $SU(5)$. We show that a suppression of the dimension five operators
 can be achieved in this case via a cancellation between  contributions
 from the $5_H+\bar 5_H$ and from the  $45_H +\overline{45}_H$.
 In Sec.~\ref{sec4} we discuss the cancellation mechanism for an $SU(5)$
  model with Planck scale contributions .
  In Sec.~\ref{sec5}, we  extend this analysis to $SO(10)$ GUT, where we consider
 the recently proposed model based on a unified Higgs sector.
Specifically,  we consider the model where the Higgs sector
consists of the SO(10) irreducible  representations
$144_H+\overline{144}_H$, which allow one  to break the $SO(10)$
gauge  group all the way down to $SU(3)_C\times U(1)_{em}$. The
$SU(5)$ decomposition of  $144_H(\overline{144}_H)$ contains
$\overline{45}_H(45_H)$ of  $SU(5)$ Higgs representations. Thus in
this case a mechanism similar to that of Sec.~\ref{sec3} for the
cancellation of baryon and lepton number  violating dimension five
operators can also be implemented.  While the analyses in
Secs.~\ref{sec3}- \ref{sec5} are  for specific  models, the
cancellation  mechanism for the suppression of baryon and lepton
number  violating dimension five operators may be more general and
applicable to a larger class of models. Conclusions are given in
Sec.~\ref{sec6}.
\section{\label{sec2}SUPPRESSION OF  HIGGSINO MEDIATED PROTON DECAY IN GUTS AND STRINGS}
In this section we consider  the constraints that are necessary
for the suppression or complete elimination of all baryon and
lepton number  violating dimension five  operators in grand
unified or  in string theory models.  Thus in such models below
the unification scale for grand unified theory or below the
string  scale for string theory the Standard Model gauge
group invariance, i.e., invariance under $SU(3)_C\times
SU(2)\times U(1)_Y$ prevails.   For the sake of the analysis below
we
 assume that the doublet-triplet problem is  resolved with one pair of Higgs
doublets light and all the remaining  Higgs doublets and all the
Higgs triplets   are heavy.  Now  grand unified theories and string
theories in general will generate Higgs triplets (anti-triplets)
with charges $Q=-1/3(1/3)$ and $Q=-4/3(4/3)$. We denote the Higgs
triplets (anti-triplets) with charges $Q=-1/3(1/3)$ by
$H^{\alpha}_q(H'_{q\alpha})$ ($q=1,2,..,n$) and $\alpha=1,2,3$ is
the color index, and Higgs triplets (anti-triplets) with charges
$Q= - 4/3(4/3)$ by $\tilde H^{\alpha}_{q'}(\tilde H'_{q'\alpha})$
($q'=1,2,..,m$). The $SU(3)_C\times SU(2)_L\times U(1)_Y$
invariant superpotential below the unification scale  may then be
written as
    \beqn (H'_{q\alpha} {\cal M}_{qp} H_p^{\alpha}
+J_{q\alpha} H^{\alpha}_q + H'_{q\alpha} K_q^{\alpha})\nonumber\\
 + (\tilde H'_{q'\alpha} \tilde {\cal M}_{q'p'} \tilde H_{p'}^{\alpha} +\tilde J_{q'\alpha} \tilde H^{\alpha}_{q'} +
 \tilde H'_{q'\alpha} \tilde K_{q'}^{\alpha}).
\eeqn For the matter content of MSSM, with three generations of
quarks and leptons, the sources  $J$ and $K$ have the following
form
 \beqn
 J_{q\alpha}=   f_{q\a\b}^{(1)}  \epsilon_{\alpha\beta\gamma} Q^{\beta}_{\a} Q^{\gamma}_{\b}
 + f_{q\a\b}^{(2)}U^C_{\a \alpha} E^C_{\b}\nonumber\\
 K_p^{\alpha} = f^{(1)'}_{p\a\b}  Q^{\alpha}_{\a}L_{\b}
 +    f^{(2)'}_{p\a\b} \epsilon^{\alpha\beta\gamma}
 U_{\a\beta}^C  D^C_{\b\gamma}.
\label{fa}
  \eeqn
  Here $Q_{\a}$ ($L_{\b}$) are quark(lepton)  $SU(2)_L$  doublets,
  and $U_{\a}^C, D^C_{\b}$ ($E^C_{\a}$) are $SU(2)_L$ singlets,
  where $\a,\b =1,2,3$ are the generation indices.  For the tilde sources
   $\tilde J$ and $\tilde K$ one  has the form
 \beqn
 \tilde J_{q'\alpha}=
 \tilde f_{q'\a\b}
 D^C_{\a \alpha} E^C_{\b},
 ~~\tilde K_{p'}^{\alpha} =
  \tilde  f'_{p'\a\b} \epsilon^{\alpha\beta\gamma}
 U_{\a\beta}^C  U^C_{\b\gamma}.
 \label{fb}
  \eeqn
Now  suppose  we make unitary transformation and go to a basis
where only $H_1$ and $H'_1$ (in the new  basis) couple  with the
matter fields. Then it is easily seen that the condition that
kills the baryon and lepton number  violating  dimension five
operators of $LLLL$ type
 is  as follows
\beqn (U_{\a\b}^{(1)}{\cal
M}V^{(1)T}_{\acute{c}\acute{d}})^{-1}_{11} +
\Lambda_{\a\b\acute{c}\acute{d}}^{QG} =0,
\label{LL} \eeqn
while for the suppression of baryon and lepton number  violating
interactions of type RRRR one has the constraint \beqn
(U_{\a\b}^{(2)}{\cal M}V^{(2)T}_{\acute{c}\acute{d}})^{-1}_{11} +
(\tilde U_{\a\b}\tilde {\cal M}\tilde
V^T_{\acute{c}\acute{d}})^{-1}_{11} +\tilde
\Lambda_{\a\b\acute{c}\acute{d}}^{QG}=0,
 \label{RR} \eeqn
Here
$U$ and $V$, and $\tilde U$ and $\tilde V$  are unitary matrices
that  take  us to the basis where only $H_1$ and $H'_1$ couple
with matter. We note that the matrices ${\cal M}$ and $\tilde {\cal M}$ 
as well as the sources $J, K, \tilde J, \tilde K$ may contain Planck scale 
contributions as exhibited explicitly in Sec.(IV). However, in  addition one
may have  quantum gravity (QG)  corrections  which we have exhibited  by
$\Lambda^{QG}$ and $\tilde \Lambda^{QG}$ terms in Eqs. (4) and  (5) 
to take account of such
effects. In this work we do not consider the quantum gravity corrections
 although such corrections could also be  utilized  for the suppression of 
 B\&L violating dimension five operators6
in grand unified and string
theory models. If we are already in the basis where only $H_1$ and
$H'_1$ couple with matter, then $U^{(1)}=I=V^{(1)}$,
$U^{(2)}=I=V^{(2)}$, and $\tilde  U=I=\tilde V$, and if we ignore
the quantum gravity  effects then one gets
  the familiar condition\cite{Arnowitt:1993pd}  ${\cal M}^{-1}_{11}=0$
 when only the Higgs  triplets (anti-triplets) with charges  $Q=-1/3(1/3)$  are considered in the
 analysis.
The constraints of Eqs.(\ref{LL}) and (\ref{RR}) together are
then sufficient
 to  kill all baryon and lepton number  violating  dimension five operators
in any grand unified theory or in any string theory model arising
from Higgsino exchange. The constraints of Eqs.(\ref{LL}) and
(\ref{RR}) are very stringent because of their dependence on
generation indices. However, significant simplification will occur
in specific
 unified models. Below we discuss two models, one in SU(5) and the other
 in SO(10) where the  constraints of   Eqs.(\ref{LL}) and (\ref{RR}) can be satisfied by
 internal cancellations.
  In the analysis below we  consider  cases where proton decay is suppressed via the cancellation
 mechanism both in the absence of the Planck scale contributions  (Sec.III) as well as when Planck
 scale contributions   are taken into account (Secs IV and V).

\section{\label{sec3}THE CANCELLATION MECHANISM IN SU(5) GRAND UNIFICATION}
In this  section we illustrate the satisfaction of  Eqs.(\ref{LL})
and (\ref{RR}) in the context of an  $SU(5)$ model. The Higgs
sector of the minimal $SU(5)$ model consists  of a $24_H$ of Higgs to
break the GUT symmetry and a pair of $5_H+\bar 5_H$ to break the
electro-weak symmetry.  Typically in this model a  fine tuning is
needed to obtain the doublet-triplet splitting. We expand now the
Higgs  sector by inclusion of  a pair of  $45_H+\overline{45}_H$
of Higgs (for the use of $45$ plet in $SU(5)$ model building see
\cite{45plet}).
 In this case the  superpotential for the Yukawa
couplings is of the form \beqn W_Y&=& f_{1\acute{a}\acute{b}}
10_{\acute{a}}.10_{\acute{b}}.5_H + f'_{1\acute{a}\acute{b}}
10_{\acute{a}}.\bar 5_{\acute{b}}.\bar 5_H\nonumber\\
 &&+
f_{2\acute{a}\acute{b}} 10_{\acute{a}}.10_{\acute{b}}.45_H+
f'_{2\acute{a}\acute{b}} 10_{\acute{a}}.\bar
5_{\acute{b}}.\overline{45}_H \label{su5yuk}.
\label{fc}
  \eeqn
For the Higgs  superpotential we choose
\beqn
W_H&=&   M_5 \bar 5_H 5_H +
h_1 \bar
5_H.24_H.5_H +  h_2 \bar 5_H.24_H.45_H +  h_3
5_H.24_H.\overline{45}_H \nonumber\\
&&+ h_4 {\overline M} 45_H.\overline{45}_H + h' W'_H(24_H).
\label{higgs} \eeqn
Here $W'_H(24_H)$  generates spontaneous
breaking producing a VEV of the form
 \beqn <24_H>=diag (2,2,2, -3,
-3) M,
\label{su5breaking}
\eeqn
and breaks $SU(5)\to SU(3)\times SU(2)\times U(1)_Y$
and we assume that there is no VEV growth for the $45$ plet of
Higgs.
In the above we adopt the  fine tuning that is
conventionally used  to produce a light Higgs .
We exhibit this explicitly.  The Higgs  doublets arise from both $5_H+\bar 5_H$
and from $45+\overline{45}_H$ and we denote  these by
    $H^{a}(5_H)$,  $P^{a}(45_H)$,
 and by $H'_{a} (\bar 5_H)$, $Q_{a}(\overline{45}_H)$. The mass matrix is
 given by

\begin{eqnarray}
&\begin{matrix}H^{a}~~~P^{a} \end{matrix}& \nonumber\\
\begin{matrix}H'_{a}\cr Q_{a}\end{matrix}
&\left(\begin{matrix}
 M^d_{11} &M^d_{12}\cr
 M^d_{21} & M^d_{22}
 \end{matrix} \right)&,
  \label{doubletsmatrix}
    \end{eqnarray}
  where for the model of Eq.(\ref{higgs})
    \beqn
  &M^d_{11}= -3Mh_1+M_5,~~M^d_{12}= -\frac{5\sqrt 3}{2\sqrt 2} Mh_2, ~~ M^d_{21}= -\frac{5\sqrt 3}{2\sqrt 2} Mh_3,&\nonumber\\
  &M^d_{22}= {\overline M} h_4.&
  \eeqn
  We denote the Higgs triplets by
    $H^{\alpha}(5_{\overline{144}})$, $Q^{\alpha}({5}_{144})$, $P^{\alpha}(45_{\overline{144}})$
 and the anti-triplets by $H'_{\alpha} (\bar 5_{144})$, $Q_{\alpha}(\overline{45}_{144})$,
  $P_{\alpha}(\overline{5}_{\overline{144}})$.
   Here  $H^{\alpha}$, $Q^{\alpha}$
  ($H'_{\alpha} $, $Q_{\alpha}$) have charges $-1/3(1/3)$ while
    $Q^{\alpha}$ ($P_{\alpha}$) have charges  -4/3(4/3).
       In the basis where the columns are
  $H^{\alpha}, P^{\alpha}, Q^{\alpha}$ and the rows  are $H'_{\alpha}, Q_{\alpha}, P_{\alpha}$,
  the Higgs  triplet mass matrix  has  the following form
\begin{eqnarray}
&\begin{matrix}H^{\alpha}~~~P^{\alpha}~~~Q^{\alpha} \end{matrix}& \nonumber\\
\begin{matrix}H'_{\alpha}\cr Q_{\alpha}\cr P_{\alpha}\end{matrix}
&\left(\begin{matrix}
 M_{11} &M_{12} & 0\cr
 M_{21} & M_{22} &0\cr
0 &0 & M_{33} \end{matrix} \right)&,
 \label{tripletsmatrix}
  \end{eqnarray}
  where for the model of Eq.(\ref{higgs})
    \beqn
  &M_{11}= 2Mh_1 + M_5, ~~M_{12}= -\frac{5}{\sqrt 2} Mh_2, ~~ M_{21}= -\frac{5}{\sqrt 2} Mh_3,&\nonumber\\
  &M_{22}= {\overline M} h_4, ~~
  M_{33}=  {\overline M} h_4.&
 \label{matrix}
  \eeqn
  From Eq.(\ref{doubletsmatrix}) one finds that all the doublets  are heavy and
  one needs  a fine tuning to get a light Higgs  doublet. This fine tuning condition is
  \beqn
  h_4= -\frac{25M}{4\bar M} {h_2h_3}({h_1- \frac{M_5}{3M}})^{-1}
  \label{finetune}
  \eeqn
  With this  constraint the second pair of Higgs  doublets are heavy and do not
  participate in low energy physics.  Further, with the constraint of Eq.(\ref{finetune})
  all the three Higgs triplets  given by Eq.(\ref{tripletsmatrix}) are heavy.\\

The Higgs triplet  interactions are
\beqn W_{int}&=& (J_{1\alpha}H^{\alpha} +
 J_{2\alpha} P^{\alpha}+
 H_{\alpha}'K_1^{\alpha} +   Q_{\alpha} K_2^{\alpha'}) \nonumber\\
 &&+
 (\tilde J_{\alpha} Q^{\alpha}
  + \tilde K^{\alpha}P_{\alpha} ),
\label{jk1} \eeqn where $J_1^{\alpha},~K_{1\alpha}$ etc. are the
matter currents to which the color  Higgs fields couple. We assume
that
  $J_1,~K_1$ arise from $5_H+\bar 5_H$ Higgs couplings, while
the $J_2,~ K_2$ and  $\tilde  J,~ \tilde K$ arise from
$45_H+\overline{45}_H$.
In order to satisfy the constraint of Eq.(\ref{LL}) and Eq.(\ref{RR})
we
make  specific assumptions regarding the generational dependence
of $45_H$ Higgs couplings relative to the $5_H$ Higgs couplings,
and of $\overline{45}_H$ Higgs coupling relative to the $\bar 5_H$
Higgs couplings as follows
 \beqn f_{2\a\b}= \lambda
f_{1\a\b},   ~~ f'_{2\a\b}= \lambda' f'_{1\a\b}.
\label{cancel} \eeqn In this case in the analysis of baryon and
lepton number violating dimension five operators the generational
dependence factors out and the entire left hand side of
Eq.(\ref{LL})  and Eq.(\ref{RR})
is proportional to $f_{1\a\b}
f'_{1\acute{c}\acute{d}}$. On eliminating the Higgs triplets fields one finds
lepton and baryon number violating operator of chirality $LLLL$ and  of
chirality $RRRR$.  The $LLLL$ operators  arise from the elimination of the
heavy Higgs fields  $H_{\alpha}', Q_{\alpha}; H^{\alpha}, P^{\alpha}$
with charges $\pm\frac{1}{3}$.
The cancellation condition in this case is
\beqn
\sum_{i=1}^{4} C_i^L(\lambda, \lambda') h_i' =0
\label{c2}
\eeqn
Here  $h_1'=(h_1+M_5/2M)$, $h_i'=h_i$ (i=2,3,4), and
 $C_i^L(\lambda, \lambda')$ is a polynomial of the type
$(a_L+b_L\lambda+ c_L\lambda'+d_L \lambda \lambda')$ where $a_L$ etc are numerical
co-efficients.
For the $B\&L$ violating dimension five operators of chirality $RRRR$,
one finds that  the contributions to them
 arise from the elimination of the
heavy Higgs fields  $H_{\alpha}', Q_{\alpha}; H^{\alpha}, P^{\alpha}$ as well
as from the elimination of the Higgs tripelts  $P_a, Q^a$ with charges  $\pm \frac{4}{3}$.
The cancellation condition in this case is
\beqn
\sum_{i=1}^{4}  C_i^R(\lambda, \lambda') \frac{\bar M}{M} h_4 h_i'
-\lambda \lambda' det ({\cal H})=0
\label{c3}
\eeqn
where $det({\cal H}) =(2\frac{\bar M}{M} h_1'h_4 -\frac{25}{2} h_2h_3)$ and
where $C_i^R$ are defined analogous to $C_i^L$.
Eqs (\ref{finetune}),(\ref{c2}),(\ref{c3})
constitute  three constraints on four
Higgs couplings $h_i$ (i=1-4) and thus can be satisfied leaving one
parameter still arbitrary.  Specifically, there are no constraints aside from
the parallelity  condition of Eq.(\ref{cancel})
 on the matter  couplings  of the Higgs which
can thus be used to fix the textures.

\section{\label{sec4}The cancellation mechanism in an $SU(5)$ model with Planck scale contributions }
In Secs (III)  we  have given an explicit demonstration of an $SU(5)$ model where  the
cancellation mechanism leads  to a  suppression of proton decay.  The analysis of Sec.III,
however, did not have any Planck scale contributions .
In this section we give a further example of the
cancellation mechanism in the context of an $SU(5)$ model including Planck scale contributions .
Such Planck scale contributions  are used to generate  the hierachical stuctures for the
 quark-lepton textures. Thus in the
notation of Sec.III we may write  the effective superpotential at the GUT scale
 including Planck scale
corrections in the form
\beqn
W=  \sum_n   \left(  f_{1n}  5.10.5_H \frac{\Sigma^n}{M_{Pl}^n} +  f_{2n}10.10.5_H    \frac{\Sigma^n}{M_{Pl}^n}  \right)
\eeqn
  where the couplings  are written in a schematic form.  After spontaneous breaking $\Sigma$ develops
  a VEV of size $M$ (see Eq. (\ref{su5breaking}))
  and the above contribute terms suppressed by powers of $(M/M_{Pl})^n$.
 % However,  some of the terms can be significantly enhanced
 % when the VEV of a  $Tr(\Sigma^2)$ appears.
 % which gives a  factor of 30 enhancement (see Eq.(\ref{su5breaking})).
% Taking such effects into account, 
An analysis of the quark-lepton textures
  using expansions  up to $(\Sigma /M_{Pl})^3$ was carried out in \cite{textures}.
    The above analysis has been extended recently
     to include  expansions  up to $(\Sigma /M_{Pl})^4$\cite{Haba:2006qp}.
        The co-efficients $a_{5L}(a_{5R})$  of the effective B\&L violating
        dimension 5 operators  $LLLL$ ($RRRR$)   take the form
%
%   \beqn
%   a_{5L}^{ijkl}= f_{ql}^{ij} f_{qq}^{kl}, ~~ a_{5R}^{ijkl}=    f_{ud}^{ij} f_{eo}^{kl}
%   \label{dim5op}
%   \eeqn
%
  \beqn
  a_{5L}^{\a\b\acute{c}\acute{d}}   =   f_{q\a\b}^{(1)}  f^{(1)'}_{p\acute{c}\acute{d}}, ~~a_{5R}^{\a\b\acute{c}\acute{d}}= f_{q\a\b}^{(2)}  f^{(2)'}_{p\acute{c}\acute{d}},
  \eeqn
  where $f_{q\a\b}^{(1)},  f^{(1)'}_{p\acute{c}\acute{d}}$ etc are as defined by Eq.(\ref{fa}).  Specifically  $f_{q\a\b}^{(1)},  f^{(1)'}_{p\acute{c}\acute{d}}$ etc  are effective
  couplings which are expansions in the Planck scale contributions . Their  forms are explicitly exhibited  in  \cite{Haba:2006qp}.
       With the larger  number of couplings available  it is
     then possible to satisfy all the quark lepton textures. Further,  one finds that solutions allow for the possibility that\cite{Haba:2006qp}
     $ f^{(1)'}_{p\acute{c}\acute{d}}=0=     f^{(2)'}_{p\acute{c}\acute{d}}$
     which completely
 suppress the B\&L violating  dimension five operators. This is an
  example of the cancellation mechanism where Planck scale   corrections allow for  the suppression of proton decay.

% The above is  another way of implementing
% the cancellation mechanism consistent with the  generation of desired quark lepton textures.

\section{SUPPRESSION OF BARYON AND LEPTON NUMBER VIOLATING DIMENSION FIVE OPERATORS IN AN SO(10) MODEL\label{sec5}}

In the $SU(5)$ model of
Sec.III  we saw that if there are  more than one pair
of Higgs  triplets contributing to the generation of baryon and
lepton number  violating interactions,   then there is  the
possibility of a partial or complete cancellation of these
operators.
In Sec,IV we saw the phenomenon of cancellation an $SU(5)$ model
with Planck scale contributions .
%Thus in an SU(5) model, the cancellation mechanism can
%work when couplings to quarks and leptons involve more than one
%pair of Higgs triplets and anti-triplets. This comes about  quite
%naturally when one has a Higgs sector involving $5_H+\bar 5_H$ and
%$45_H + \overline{45}_H$.
We consider  now  the $SO(10)$ case.
There is already a considerable literature  on model building in
SO(10) (for some recent works see\cite{pdecay3,pdecay4,Dutta:2004zh}).
Here we will consider the $SO(10)$ model proposed recently
 with one step breaking
down to the Standard Model gauge group and further down to the residual
gauge group $SU(3)_C\times U(1)_Y$ using $144_H+\overline{144}_H$  of Higgs\cite{bgns1}.
%This case is different from the analysis of Sec.III  in that additionally
%Planck scale contributions   play an essential role in this case.
This case  combines  some features of the models discussed in Sec.III and
in Sec.IV. Thus the model has more than one pair of Higgs triplets, and
further,  it has  Planck scale contributions .
Thus the quark-lepton
masses for the first  two generations (see  Sec(\ref{quarticcoup}) )  arise from the Planck scale contributions ,
while that of the third generation arise  from the cubic interactions (see Eq.(\ref{cubic1})).
 We will compute the lepton and baryon number violating interactions for  this model
 and show that a complete suppression of baryon and lepton number  violating dimension 5 operators can occur in
 this case.

We begin by exhibiting the decomposition of $144$ under
$SU(5)\times U(1)$. Here one finds \beqn\label{144} 144&=& 5 ({\bf
Q}^i) [3]+\bar 5 ({\bf Q}_i) [7]+ 10 ({\bf Q}^{ij})[-1] + 15 ({\bf
Q}_{(S)}^{ij}) [-1]\nonumber\\
&&+ 24({\bf Q}^i_j)[-5] + 40({\bf Q}^{ijk}_l)[-1] +
\overline{45}({\bf Q}^{i}_{jk})[3], \eeqn
where $i,j,k$ are the SU(5) indices and a similar decomposition of
$\overline{144}$ holds so that we have
 \beqn  \label{bar144}
 \overline{144} &=&\bar 5
({\bf P}_i)[-3]+5 ({\bf P}^i)[-7]+ \overline{10}({\bf P}_{ij})[1]
+ \overline{15}({\bf P}^{(S)}_{ij})[1] \nonumber\\
&&+ 24 ({\bf P}^i_j)[5] + \overline{40}({\bf P}^l_{ijk})[1] +
45({\bf P}^{ij}_{k})[-3].
 \eeqn
 We note that the decomposition of $144_H+\overline{144}_H$
contains  $5+\bar 5$ pairs of Higgs, as well as a pair of
$45_H+\overline{45}_H$  of Higgs. Thus in this sense it contains
the essential ingredients of the $SU(5)$ model which has $5+\bar
5$ and $45_H+\overline{45}_H$ of Higgs fields. There is then a
good chance that a cancellation mechanism works in this case as
well.   We will show later in this section that this is indeed the
case.
 The  analysis in the rest of this section is as follows: in Sec.~(VA) we give a  brief discussion  of
 spontaneous  breaking with $144_H+\overline{144}_H$   of Higgs. In Sec.~(VB) we discuss
 the doublet -triplet splitting.
 In Sec.~(VC) we give an $SU(5)\times U(1)$
  decomposition of  couplings of matter and Higgs. Here we analyze
 quartic interactions  as well as cubic interactions\cite{bgns2}     when
 additional $10$ and $45$ of matter are introduced  in  order to generate large
 masses for the third generation of quarks and leptons.  An analysis of baryon and lepton number  violating
 interactions is
 given in Sec.~(VD) where the condition for the complete suppression of baryon and lepton number
 violating  LLLL and RRRR operators by
 the cancellation mechanism is discussed.

\subsection{\label{sec41}Spontaneous symmetry breaking}
To discuss the spontaneous breaking with $144_H+\overline{144}_H$
of Higgs, we consider the following form for the superpotential
\begin{eqnarray}
 {\mathsf W}&=& M {(\overline{144}_H\times 144_H)}\nonumber\\
 &&+ \frac{\lambda_{45_1}}{M'} { (\overline{144}_H\times
144_H)_{45_1}
 (\overline{144}_H\times 144_H)_{45_1}}\nonumber\\
 &&+  \frac{\lambda_{45_2}}{M'} { (\overline{144}_H\times
144_H)_{45_2}
 (\overline{144}_H\times 144_H)_{45_2}}\nonumber\\
 &&+  \frac{\lambda_{210}}{M'} { (\overline{144}_H\times
144_H)_{210}
 (\overline{144}_H\times 144_H)_{210}}.\nonumber\\
 \label{quartic}
\end{eqnarray}
In the above the $45_1$, $45_2$ and $210$ couplings  are defined as follows
\begin{eqnarray}
 (\overline{144}_H\times 144_H)_{45_1}
 (\overline{144}_H\times 144_H)_{45_1}=   <\Psi^{*}_{(-)\mu}|B\Sigma_{\rho\lambda}|\Psi_{(+)\mu}>
 <\Psi^{*}_{(-)\nu}|B\Sigma_{\rho\lambda}|\Psi_{(+)\nu}>\\
 (\overline{144}_H\times 144_H)_{45_2}
 (\overline{144}_H\times 144_H)_{45_2}=    <\Psi^{*}_{(-)[\mu}|B|\Psi_{(+)\nu]}>
 <\Psi^{*}_{(-)[\mu}|B|\Psi_{(+)\nu]}>\\
 (\overline{144}_H\times 144_H)_{210}
 (\overline{144}_H\times 144_H)_{210}=
 <\Psi^{*}_{(-)\mu}|B\Gamma_{[\rho}
 \Gamma_{\sigma}\Gamma_{\lambda}\Gamma_{\xi
 ]}|\Psi_{(+)\mu}>\nonumber\\
\cdot<\Psi^{*}_{(-)\nu}|B\Gamma_{[\rho}
 \Gamma_{\sigma}\Gamma_{\lambda}\Gamma_{\xi ]}|\Psi_{(+)\nu}>
\end{eqnarray}
In the above $\Gamma_{\mu}$ ($\mu=1,2,...,10$)
 are the $SO(10)$ matrices which satisfy the Clifford  algebra
 \beqn
 \{\Gamma_{\mu}, \Gamma_{\nu}\} = 2\delta_{\mu\nu}
 \eeqn
and $B$ is the $SO(10)$ charge conjugation matrix
\beqn
B= \prod_{\mu ={\rm odd}} \Gamma_{\mu}
\eeqn
The explicit computations are done using the oscillator
method\cite{ms,wilczek} and the techniques developed in
\cite{bgns1,ns,ns1}.  These techniques are field theoretic and the
$144_H(\overline{144}_H)$  plet in this scheme is  represented by
a constrained vector spinors $|\Upsilon_{(\pm)\mu}>$ where one
imposes the
 constraint
  $\Gamma_{\mu}
|\Upsilon_{(\pm)\mu}> =0$.
%Here $\Gamma_{\mu}$  are $SO(10)$ gamma
%matrices which satisfy a rank-10 Clifford algebra
%$\{\Gamma_{\mu},\Gamma_{\nu}\}=2\delta_{\mu\nu}$,
% where $\mu,\nu$ are the $SO(10)$ indices and take on the
%values $1,..,10$.
We carry out an  explicit analysis and find
\begin{eqnarray}
{\mathsf W}_{_{SB}}=M{ \bf Q}^i_{j}{\bf P}^j_{i}
+\frac{1}{M'}\left[-\lambda_{45_{_1}}+\frac{1}{6}\lambda_{210}
\right] {
\bf Q}^i_{j}{ \bf P}^j_{i} {\bf Q}^k_{l}{ \bf P}^l_{k}\nonumber\\
 +\frac{1}{M'}\left[-4\lambda_{45_{_1}}-\frac{1}{2}\lambda_{45_{_2}}-\lambda_{210}
\right] {\bf Q}^i_{k}{ \bf P}^k_{j} {\bf Q}^j_{l}{ \bf P}^l_{i},\nonumber\\
\end{eqnarray}
The minimization of ${\mathsf W}_{_{SB}}$ gives
\begin{eqnarray}
 <{\bf Q}^i_{j}>&=& q~\textnormal{diag}(2,2,2,-3,-3)\nonumber\\
 <{\bf P}^i_{j}>&=& p~\textnormal{diag}(2,2,2,-3,-3),
\label{pdef}
\end{eqnarray}
where $q,p$ are constrained by
\begin{eqnarray}\label{symmetrybreaking}
\frac{MM'}{qp}=116\lambda_{45_{_1}}+7\lambda_{45_{_2}}
+4\lambda_{210}.
\end{eqnarray}
\subsection{\label{sec42}Doublet-triplet splitting}
We discuss now the doublet-triplet splitting.
 Our general philosophy is that we are working within the context of a landscape scenario where a fine tuning
 to make the Higgs  doublet light is permissible\cite{Arkani-Hamed:2004fb,Kors:2004hz}. After spontaneous breaking of the electroweak symmetry
   discussed in the preceding section, one finds that
 the part of the  superpotential  that governs the doublet-triplet splitting is given by
\begin{widetext}
\begin{eqnarray}
 {\mathsf
 W}_{_{DT}}&=&\left\{\frac{4}{5}M+\frac{1}{M'}\left(\frac{24}{5}\lambda_{45_{_1}}
 -\frac{4}{15}\lambda_{210}\right)<{\bf Q}^m_{n}><{\bf P}^n_{m}>\right\}
 {\bf Q}_{i}{\bf P}^i\nonumber\\
&&+\left\{\frac{1}{M'}\left[-\frac{4}{5}\lambda_{45_{_2}}
-\frac{32}{15} \lambda_{210}\right]<{\bf Q}^m_{j}><{\bf
P}^i_{m}>\right\}
{\bf Q}_{i}{\bf P}^j\nonumber\\
&&+\left\{M+\frac{1}{M'}\left(6\lambda_{45_{_1}}
 -\frac{1}{3}\lambda_{210}\right)<{\bf Q}^m_{n}><{\bf P}^n_{m}>\right\}
 {\bf Q}^i{\bf P}_{i}\nonumber\\
&&+\left\{\frac{1}{M'}\left(\lambda_{45_{_2}} \right) <{\bf
Q}^m_{i}><{\bf P}^j_{m}>\right\}
{\bf Q}^i_{}{\bf P}_{j}\nonumber\\
&&+\left\{-\frac{1}{2}M+\frac{1}{M'}\left(\lambda_{45_{_1}}
-\frac{1}{6}\lambda_{210}\right)<{\bf Q}^m_{n}><{\bf
P}^n_{m}>\right\} \left[{\bf Q}_{ij}^k+\frac{1}{2\sqrt
5}\left(\delta^k_i{\bf Q}_{j}-\delta^k_j{\bf Q}_{i}\right)\right]\nonumber\\
&&\times \left[{\bf P}^{ij}_{k}+\frac{1}{2\sqrt
5}\left(\delta^i_k{\bf P}^j-\delta^j_k{\bf P}^i\right)\right]
\nonumber\\
&&+\left\{\frac{1}{M'}\left(-\frac{1}{2}\lambda_{45_{_2}} \right)
<{\bf Q}^m_{i}><{\bf P}^j_{m}>\right\} \left[{\bf Q}_{kl}^i+\frac{1}{2\sqrt 5}\left(\delta^i_k{\bf Q}_{l}-\delta^i_l{\bf Q}_{k}\right)\right]\nonumber\\
&&\times \left[{\bf P}^{kl}_{j}+\frac{1}{2\sqrt
5}\left(\delta^k_j{\bf P}^l-\delta^l_j{\bf P}^k\right)\right]
\nonumber\\
&&+\left\{\frac{1}{M'}\left[\left(8\lambda_{45_{_1}}-\frac{2}{3}\lambda_{210}
\right) <{\bf Q}^i_{m}><{\bf
P}^m_{j}>\right]\right\}\left[{\bf Q}_{il}^k+\frac{1}{2\sqrt 5}\left(\delta^k_i{\bf Q}_{l}-\delta^k_l{\bf Q}_{i}\right)\right]\nonumber\\
&& \times \left[{\bf P}^{lj}_{k}+\frac{1}{2\sqrt
5}\left(\delta^l_k{\bf P}^j-\delta^j_k{\bf P}^l\right)\right].
\end{eqnarray}
\end{widetext}
In the limit when $\lambda_{45_1}=0=\lambda_{45_2},
\lambda_{210}=0$, the Higgs doublets and triplets that  pair up
 are the following set
\begin{eqnarray}
&{\mathsf D}_1:~({\bf Q}^{a}, {\bf P}_{a});~~~~~~~~{\mathsf T}_1:~({\bf Q}^{\alpha}, {\bf P}_{\alpha})&\nonumber\\
&{\mathsf D}_2:~({\bf Q}_{a},{\bf P}^{a} );~~~~~~~~{\mathsf T}_2:~({\bf Q}_{\alpha},{\bf P}^{\alpha})&\nonumber\\
&{\mathsf D}_3:~({\bf{\widetilde Q}}_{a}, {\bf {\widetilde P}}^{ a
});~~~~~~~~{\mathsf T}_3:~({\bf{\widetilde Q}}_{\alpha}, {\bf
{\widetilde P}}^{ \alpha
})&\nonumber\\
&{\mathsf T}_4:~({\bf{\widetilde Q}}^{\alpha}, {\bf {\widetilde
P}}_{ \alpha }).&
\end{eqnarray}
Here  ${\mathsf  T}_1$, ${\mathsf  T}_2$, ${\mathsf  T}_3$ are
color triplet (anti-triplet) pairs which have  charges
$Q=-1/3(1/3)$, while ${\mathsf  T}_4$ have charges $Q=-4/3(4/3)$.
Thus the pattern of color Higgs multiplets discussed in
Sec.~\ref{sec3} is reproduced here.

Mixings among the mutiplets occur when the couplings
$\lambda_{45_1}, \lambda_{45_2}, \lambda_{210}$ are non-zero. For
the Higgs  doublet fields the doublet  ${\mathsf  D}_1$ is decoupled while the
 mixings  that lead to the doublets  ${\mathsf  D}_2$ and  ${\mathsf  D}_3$  are generated by the  mass matrix
 \begin{widetext}
{ \beqn &\begin{matrix}{\bf
Q}_{a}~~~~~~~~~~~~~~~~~~~~~~~~~~~~~~~~~~~~~~~~~~~~~~~~~ {\bf
{\widetilde Q}}_{a}  \end{matrix}&
\nonumber\\
\begin{matrix}{\bf P}^{a}\cr\cr \bf {\widetilde P}^{a}\end{matrix}
&\left[\begin{matrix}
\frac{3}{5}M+\frac{qp}{M'}(\frac{666}{5}\lambda_{45_1}-\frac{33}{4}\lambda_{45_2}
-\frac{273}{10}\lambda_{210})
&\sqrt{\frac{3}{5}}\frac{qp}{M'}(10\lambda_{45_1}+\frac{5}{4}\lambda_{45_2}
-\frac{5}{6}\lambda_{210})  \cr\cr
 \sqrt{\frac{3}{5}}\frac{qp}{M'}(10\lambda_{45_1}+\frac{5}{4}\lambda_{45_2}
-\frac{5}{6}\lambda_{210}) &
 -\frac{1}{2}M+\frac{qp}{M'}(-74\lambda_{45_1}-\frac{31}{4}\lambda_{45_2}+\frac{7}{6}\lambda_{210})
 \end{matrix}\right]&.~~~
\eeqn }
  For the triplets, ${\mathsf  T}_1$ and ${\mathsf  T}_4$ are decoupled  while  
  ${\mathsf  T}_2$ and ${\mathsf  T}_3$ mix. 
   % For the triplets,  $T_1$ and $T_4$ are decoupled while
%$T_2$ and $T_3$ mix.
 The mixings that  lead to the  Higgs triplets
%$T_2$ and $T_3$ 
 ${\mathsf  T}_2$ and ${\mathsf  T}_3$ 
are given by the mass matrix 
 {%\footnotesize
 { \beqn
&\begin{matrix}{\bf
Q}_{\alpha}~~~~~~~~~~~~~~~~~~~~~~~~~~~~~~~~~~~~~~~~~~~~~~~~~{\bf
{\widetilde
Q}}_{\alpha}  \end{matrix} \nonumber\\
\begin{matrix}{\bf P}^{\alpha}\cr\cr \bf {\widetilde P}^{\alpha}
\end{matrix}
&\left[\begin{matrix}
\frac{3}{5}M+\frac{qp}{M'}(\frac{696}{5}\lambda_{45_1}-\frac{9}{2}\lambda_{45_2}
-\frac{257}{15}\lambda_{210}) &\frac{\sqrt
5}{2}\frac{qp}{M'}(8\lambda_{45_1}+\lambda_{45_2}
-\frac{2}{3}\lambda_{210})  \cr\cr
 \frac{\sqrt 5}{2}\frac{qp}{M'}(8\lambda_{45_1}+\lambda_{45_2}
-\frac{2}{3}\lambda_{210}) &
 -M+\frac{qp}{M'}(-24\lambda_{45_1}-\frac{13}{2}\lambda_{45_2}-3\lambda_{210})
\cr\cr
\end{matrix}\right]&. ~~~ \eeqn }}
\end{widetext}
We represent  the mass eigenstates  by primed fields, and  the
primed  fields  may be expressed in terms of the unprimed  ones
through the following transformation matrices
\begin{eqnarray}\label{rotatedhiggs}
 \left[\begin{matrix}({\bf { Q}}^{\prime
}_{a},{\bf { P}}^{\prime a} )\cr ({\bf {\widetilde
Q}}^{\prime}_{a},{\bf {\widetilde P}}^{\prime a})
\end{matrix}\right]& =& \left[\begin{matrix} \cos\vartheta_{\mathsf
D} & \sin\vartheta_{\mathsf D}\cr -\sin\vartheta_{\mathsf D} &
\cos\vartheta_{\mathsf
D}\end{matrix}\right]\left[\begin{matrix}({\bf{ Q}}_{a}, {\bf {
P}}^{a})\cr ({\bf {\widetilde Q}}_{ a},{\bf {\widetilde P}}^{ a})
 \end{matrix}\right]\nonumber\\
 \nonumber\\
 \left[\begin{matrix}({\bf { Q}}^{\prime
}_{\alpha},{\bf { P}}^{\prime \alpha} )\cr ({\bf {\widetilde
Q}}^{\prime}_{\alpha},{\bf {\widetilde P}}^{\prime \alpha})
\end{matrix}\right]& =& \left[\begin{matrix} \cos\vartheta_{\mathsf T} &
\sin\vartheta_{\mathsf T}\cr -\sin\vartheta_{\mathsf T} &
\cos\vartheta_{\mathsf
T}\end{matrix}\right]\left[\begin{matrix}({\bf{ Q}}_{\alpha}, {\bf
{ P}}^{\alpha})\cr ({\bf {\widetilde Q}}_{ \alpha},{\bf
{\widetilde P}}^{ \alpha})
 \end{matrix}\right],
\end{eqnarray}
where
\begin{eqnarray}
\tan\vartheta_{\mathsf D}&=&\frac{1}{{\mathsf d_3}}\left({\mathsf
d_2}+\sqrt{{\mathsf d_2}^2+{\mathsf d_3}^2}\right)\nonumber\\
\tan\vartheta_{\mathsf T}&=&\frac{1}{{\mathsf t_3}}\left({\mathsf
t_2}+\sqrt{{\mathsf t_2}^2+{\mathsf t_3}^2}\right),
\end{eqnarray}
and where
\begin{eqnarray}
{\mathsf
d}_1&=&-\frac{2}{5}M+\frac{qp}{M'}\left(\frac{296}{5}\lambda_{45_1}
-16\lambda_{45_2}-\frac{392}{15}\lambda_{210}\right)\nonumber\\
{\mathsf
d}_2&=&-\frac{8}{5}M+\frac{qp}{M'}\left(-\frac{1036}{5}\lambda_{45_1}
+\frac{1}{2}\lambda_{45_2}+\frac{427}{15}\lambda_{210}\right)\nonumber\\
{\mathsf
d_3}&=&2\sqrt{\frac{3}{5}}\frac{qp}{M'}\left(10\lambda_{45_1}
+\frac{5}{4}\lambda_{45_2}-\frac{5}{6}\lambda_{210}\right),
\end{eqnarray}
\begin{eqnarray}
{\mathsf
t}_1&=&-\frac{2}{5}M+\frac{qp}{M'}\left(\frac{576}{5}\lambda_{45_1}
-11\lambda_{45_2}-\frac{302}{15}\lambda_{210}\right)\nonumber\\
{\mathsf
t}_2&=&-\frac{8}{5}M+\frac{qp}{M'}\left(-\frac{816}{5}\lambda_{45_1}
-2\lambda_{45_2}+\frac{212}{15}\lambda_{210}\right)\nonumber\\
{\mathsf t_3}&=&\sqrt{5}\frac{qp}{M'}\left(8\lambda_{45_1}
+\lambda_{45_2}-\frac{2}{3}\lambda_{210}\right).
\end{eqnarray}
The mass eigenvalues are found to be
\begin{eqnarray}\label{doublets}
M_{{\mathsf
D}_1}&=&M+\frac{qp}{M'}(180\lambda_{45_1}+9\lambda_{45_2}-10\lambda_{210})\nonumber\\
M_{{\mathsf D}_2,{\mathsf D}_3}&=&\frac{1}{2}\left({\mathsf
d_1}\pm\sqrt{{\mathsf d_2}^2+{\mathsf d_3}^2}\right),
\end{eqnarray}
and
\begin{eqnarray}
M_{{\mathsf
T}_1}&=&M+\frac{qp}{M'}(180\lambda_{45_1}+4\lambda_{45_2}-10\lambda_{210})\nonumber\\
M_{{\mathsf
T}_4}&=&-M+\frac{qp}{M'}(-84\lambda_{45_1}-4\lambda_{45_2}+2\lambda_{210})\nonumber\\
M_{{\mathsf T}_2,{\mathsf T}_3}&=&\frac{1}{2}\left({\mathsf
t_1}\pm\sqrt{{\mathsf t_2}^2+{\mathsf t_3}^2}\right).
\end{eqnarray}
The above allow for making one pair of Higgs doublets light by a
constraint  while all the Higgs triplets remain heavy.

\subsection{\label{sec43}Matter-Higgs interactions}
The 16-plet of matter can interact with 144-plet of Higgs only via
quartic couplings. Here we consider the following interactions
\[
\begin{array} {c}
~\left\{\zeta_{\acute{a}\acute{b},\acute{c}\acute{d}}^{^{(10)(+)}}\right\}~\left(16_{\acute{a}}\times
16_{\acute{b}}\right)_{10}\left(144_{\acute{c}}\times
144_{\acute{d}}\right)_{10}\nonumber\\
~\left\{\xi_{\acute{a}\acute{b},\acute{c}\acute{d}}^{^{(10)(+)}}\right\}~\left(16_{\acute{a}}\times
16_{\acute{b}}\right)_{10}\left(\overline{144}_{\acute{c}}
\times \overline{144}_{\acute{d}}\right)_{10}\nonumber\\
~\left\{\varrho_{\acute{a}\acute{b},\acute{c}\acute{d}}^{^{(126,\overline{126})(+)}}\right\}~\left(16_{\acute{a}}\times
16_{\acute{b}}\right)_{\overline{126}} \left(144_{\acute{c}}\times
144_{\acute{d}}\right)_{126}\nonumber\\
~\left\{\lambda_{\acute{a}\acute{b},\acute{c}\acute{d}}^{^{(45)}}\right\}~\left(16_{\acute{a}}\times
\overline{144}_{\acute{b}}\right)_{45} \left(16_{\acute{c}}\times \overline{144}_{\acute{d}}\right)_{45},\nonumber\\
~\left\{\zeta_{\acute{a}\acute{b},\acute{c}\acute{d}}^{^{(120)(-)}}\right\}~\left(16_{\acute{a}}\times
16_{\acute{b}}\right)_{120}\left({144}_{\acute{c}}\times{144}_{\acute{d}}\right)_{120}\nonumber\\
~\left\{\xi_{\acute{a}\acute{b},\acute{c}\acute{d}}^{^{(120)(-)}}\right\}~\left(16_{\acute{a}}\times
16_{\acute{b}}\right)_{{120}}
\left(\overline{144}_{\acute{c}}\times \overline{144}_{\acute{d}}\right)_{120}\nonumber\\
~\left\{\lambda_{\acute{a}\acute{b},\acute{c}\acute{d}}^{^{(54)}}\right\}~\left(16_{\acute{a}}\times
\overline{144}_{\acute{b}}\right)_{54} \left(16_{\acute{c}}\times
\overline{144}_{\acute{d}}\right)_{54}\nonumber\\
~\left\{\lambda_{\acute{a}\acute{b},\acute{c}\acute{d}}^{^{(10)}}\right\}~\left(16_{\acute{a}}\times
144_{\acute{b}}\right)_{10}\left(16_{\acute{c}}\times
144_{\acute{d}}\right)_{10},
\label{quarticcoup}
\end{array}
\]
 \noindent which contribute to the masses of quarks and leptons.
  Since these couplings are quartic they are Planck scale suppressed.
We assume that the first two generation masses arise from such couplings
while the third generation masses arise from cubic interactions.
The quantities within  \{\}  are  parameters  associated with the
particular quartic couplings with which  they appear.
The matter-Higgs quartic couplings  can be decomposed in $SU(5)$
representations as  follows \beq
 W_4= \sum_{i=1}^5 W^{(i)}_4.
\eeq The explicit analysis  of the couplings in its $SU(5)\times
U(1)$ decomposed form is carrried  out using oscillator
method\cite{ms,wilczek} and the techniques developed in
\cite{bgns1,ns,ns1}.  The result of the analysis is recorded in Appendix A.
It was noted in \cite{bgns2} that much   larger masses for the
third generation can be
 obtained if one allows for the mixings of the 16 plets of matter in the third generation
 with 10 and 45 plets of matter. Thus one may have cubic couplings of the type
 \beqn
 (16.10.144_H), ~~(16.45.\overline{144}_H).
 \label{cubic}
 \eeqn
We note in passing that the  particle content of $10$ and $45$ of matter in its
$SU(2)\times SU(3)_C\times U(1)$ decomposition is as follows: for the $10$ plet of matter
we have $10=(1,1)(6)+ (1,\bar 3)(-4)+ (2,3)(1)$ while the  $45$-plet has the
decomposition $45=(2,1)(3)+(1,3)(-2)+ (3,3)(-2) +(1,\bar 3) (8) +(2,\bar 3) (-7) +(1,\bar 6)(-2)
+(2,8)(3)$.
An explicit computation of the couplings in $SU(5)\times U(1)$ decomposition
 using  the techniques of
\cite{bgns1,ns,ns1} gives
\begin{eqnarray}
{\mathsf W}^{16\times \overline{144} \times
{45}}&=&f^{(45)}_{\acute{a}\acute{b}}\left[
\frac{1}{\sqrt{10}}\epsilon_{ijklm} {\bf M}_{\acute{a}}^{ij} {\bf
P}^k { {\bf F}}_{\acute{b}}^{(45)lm} +\frac{1}{\sqrt
2}\epsilon_{ijklm} {\bf M}_{\acute{a}}^{ij} {\bf P}_{n}^{kl} {
{\bf F}}_{\acute{b}}^{(45)mn} -2\sqrt 2 {\bf M}_{\acute{ai}} {\bf
P}_j { {\bf F}}_{\acute{b}}^{(45)ij}+...\right]\nonumber\\
 {\mathsf W}^{16\times {144} \times {10}}&=& f^{(10)}_{\acute{a}\acute{b}}\left[-\frac{1}{2\sqrt{10}}
{\bf M}_{\acute{a}}^{ij} {\bf Q}_j { {\bf
F}}_{\acute{b}i}^{(10)}+\frac{1}{2\sqrt{2}}{\bf
M}_{\acute{a}}^{ij} {\bf Q}^k_{ij} { {\bf
F}}_{\acute{b}k}^{(10)}+...\right].
\label{cubic1}
\end{eqnarray}
Using the above interactions, one can generate a realistic model
of
 quark-lepton-neutrino textures. However, a detailed  analysis  of the textures
  generated by the interactions above and fits to the experimental data is outside
  the scope of this work. Here we focus on the baryon and lepton number  violating dimension five operators
generated by the interactions above and how they can be suppressed
consistent
  with the current data.
\subsection{\label{sec44}Baryon and lepton number violating dimension-5 operators}
Using Eqs.(\ref{candidates down}), (\ref{candidates up}),
 and (\ref{cubic1}) and inserting mass terms for
triplets responsible for proton decay, we find
\begin{eqnarray}\label{B-L}
{\mathsf W}_{B\&L}&=&J_{1}^{\alpha}{\bf { P}}_{ \alpha
}+K_{1{\alpha}}{\bf { Q}}^{ \alpha }+M_{{\mathsf T}_1}{\bf {
Q}}^{ \alpha }{\bf { P}}_{ \alpha }\nonumber\\
&&+\left[J_{2{\alpha}}\cos\vartheta_{\mathsf
T}+J_{3{\alpha}}\sin\vartheta_{\mathsf T}\right]{\bf { P}}^{\prime
\alpha }+\left[K_{2}^{\alpha}\cos\vartheta_{\mathsf
T}+K_{3}^{\alpha}\sin\vartheta_{\mathsf T}\right]{\bf {
Q}}_{\alpha}^{\prime  }+M_{{\mathsf T}_2}{\bf {
Q}}_{\alpha}^{\prime }{\bf { P}}^{\prime \alpha }\nonumber\\
&&+\left[-J_{2{\alpha}}\sin\vartheta_{\mathsf
T}+J_{3{\alpha}}\cos\vartheta_{\mathsf T}\right]{\bf {\widetilde
P}}^{\prime \alpha }+\left[-K_{2}^{\alpha}\sin\vartheta_{\mathsf
T}+K_{3}^{\alpha}\cos\vartheta_{\mathsf T}\right]{\bf {\widetilde
Q}}_{\alpha}^{\prime}+M_{{\mathsf T}_3}{\bf {\widetilde
Q}}_{\alpha}^{\prime}
{\bf {\widetilde P}}^{\prime \alpha }\nonumber\\
&&+J_{4}^{\alpha}{\bf {\widetilde P}}_{\alpha}^{\prime
}+K_{4{\alpha}}{\bf {\widetilde Q}}^{\prime \alpha }+M_{{\mathsf
T}_4}{\bf {\widetilde Q}}^{\prime \alpha }{\bf {\widetilde
P}}_{\alpha}^{\prime }.
\end{eqnarray}
Here we have defined
\begin{eqnarray}
J_{1}^{\alpha}&=&2p\left[4\left(4\xi_{\acute{a}\acute{b}}^{^{(10)(+)}}
-\lambda_{\acute{a},\acute{b}}^{^{(45)}}\right)\left(\epsilon^{\alpha\beta\gamma}{\bf{D}}_{L\acute
{a}\beta}^{\mathsf c}{\bf{U}}_{L\acute {b}\gamma}^{\mathsf
c}\right) +\left(-16\xi_{\acute{a}\acute{b}}^{^{(10)(+)}}
-\lambda_{\acute{a},\acute{b}}^{^{(45)}}+5\lambda_{\acute{a},\acute{b}}^{^{(54)}}\right)
\left({\bf E}_{L\acute {a}}{\bf{U}}^{\alpha}_{L\acute
{b}}+{\Nu}_{L\acute {a}}{\bf{D}}^{\alpha}_{L\acute
{b}}\right)\right]\nonumber\\
&&-2\sqrt {2}f^{(45)}_{33}
\left(-\epsilon^{\alpha\beta\gamma}~{}^{({\overline{5}}_{16})}\!{\bf
{ b}}_{{ L\beta}}^{\mathtt c}~{}^{({{10}}_{45})}\!{\bf {t}}_{{
L\gamma}}^{\mathtt c}~+~~{}^{({\overline{5}}_{16})}\!{\bTau}
_{L}~{}^{({{10}}_{45})}\!{\bf { t}}_{{
L}}^{\alpha}~+~{}^{({\overline{5}}_{16})}\!{\bNu}_{L\tau}~{}^{({{10}}_{45})}\!{
{\bf{b}}}_{{
L}}^{\alpha}\right),\nonumber\\
\nonumber\\
 K_{1\alpha}&=&
32q\left(\zeta_{\acute{a}\acute{b}}^{^{(10)(+)}}+\frac{2}{15}
\varrho_{\acute{a}\acute{b}}^{^{(126,\overline{126})(+)}}\right)\left({\bf{U}}_{L\acute
{a}\alpha}^{\mathsf c}{\bf E}^{c}_{L\acute
{b}}-\epsilon_{\alpha\beta\gamma}{\bf{U}}_{L\acute
{a}}^{\beta}{\bf{D}}_{L\acute {b}}^{\gamma}\right),\nonumber\\
\nonumber\\
J_{2\alpha}&=& \frac{4p}{\sqrt
5}\left(-4\xi_{\acute{a}\acute{b}}^{^{(10)(+)}}+
\lambda_{\acute{a},\acute{b}}^{^{(45)}}\right)\left({\bf{U}}_{L\acute
{a}\alpha}^{\mathsf c}{\bf E}^{c }_{L\acute
{b}}-\epsilon_{\alpha\beta\gamma}{\bf{U}}_{L\acute
{a}}^{\beta}{\bf{D}}_{L\acute {b}}^{\gamma}\right)\nonumber\\
&&+2\sqrt{\frac{2}{5}}f^{(45)}_{33}\left({}^{({{10}}_{16})}\!{\bf
{t}}_{{ L\alpha}}^{\mathtt c}~{}^{({{10}}_{45})}\!{\bTau}^{c
}_{L}~+~ {}^{({{10}}_{16})}\!{\bTau}^{c}_{L}~{}^{({{10}}_{45})}\!{\bf  {t}}_{{ L\alpha}}^{\mathtt
c}~+~\epsilon_{\alpha\beta\gamma}~{}^{({{10}}_{16})}\!{\bf {t}}_{{
L}}^{\beta}~{}^{({{10}}_{45})}\!{\bf {b}}_{{
L}}^{\gamma}~-~\epsilon_{\alpha\beta\gamma}~{}^{({{10}}_{16})}\!{\bf
{b}}_{{ L}}^{\beta}~{}^{({{10}}_{45})}\!{\bf  {t}}_{{ L}}^{\gamma}
\right),\nonumber\\
\nonumber\\
K_{2}^{\alpha}&=&\frac{q}{\sqrt
5}\left[\left(16\zeta_{\acute{a}\acute{b}}^{^{(10)(+)}}
+2\lambda_{\acute{a},\acute{b}}^{^{(10)}}
-\frac{3}{5}\varrho_{\acute{a}\acute{b}}^{^{(126,\overline{126})(+)}}\right)
\left(\epsilon^{\alpha\beta\gamma}{\bf{D}}_{L\acute
{a}\beta}^{\mathsf c}{\bf{U}}_{L\acute {b}\gamma}^{\mathsf
c}\right)
\right.\nonumber\\
&&\left.+\left(-16\zeta_{\acute{a}\acute{b}}^{^{(10)(+)}}
+3\lambda_{\acute{a},\acute{b}}^{^{(10)}}
-\frac{14}{15}\varrho_{\acute{a}\acute{b}}^{^{(126,\overline{126})(+)}}\right)
\left({\bf E}_{L\acute {a}}{\bf{U}}^{\alpha}_{L\acute
{b}}+{\Nu}_{L\acute {a}}{\bf{D}}^{\alpha}_{L\acute
{b}}\right)\right]\nonumber\\
&&-\frac{1}{2\sqrt{10}}f^{(10)}_{33}\left(\epsilon^{\alpha\beta\gamma}~{}^{({{10}}_{16})}\!{\bf
{t}}_{{ L\beta}}^{\mathtt c}~{}^{({\overline{5}}_{10})}\!{\bf
{b}}_{{ L\gamma}}^{\mathtt c}~+~{}^{({{10}}_{16})}\!{\bf {t}}_{{
L}}^{\alpha}~{}^{({\overline{5}}_{10})}\!{\bTau}
_{L}~+~{}^{({{10}}_{16})}\!{\bf {b}}_{{
L}}^{\alpha}~{}^{({\overline{5}}_{10})}\!{\bNu} _{L\tau}\right),\nonumber\\
\nonumber\\
J_{3\alpha}&=&
{20p}\left[\left(4\xi_{\acute{a}\acute{b}}^{^{(10)(+)}}-
\lambda_{\acute{a},\acute{b}}^{^{(45)}}\right)\left({\bf{U}}_{L\acute
{a}\alpha}^{\mathsf c}{\bf E}^{c}_{L\acute
{b}}\right)+\left(-4\xi_{\acute{a}\acute{b}}^{^{(10)(+)}}+
\lambda_{\acute{a},\acute{b}}^{^{(54)}}\right)\left(\epsilon_{\alpha\beta\gamma}{\bf{U}}_{L\acute
{a}}^{\beta}{\bf{D}}_{L\acute {b}}^{\gamma}\right)
\right]\nonumber\\
&&+2\sqrt{2} f^{(45)}_{33}\left(-~{}^{({{10}}_{16})}\!{\bf {t}}_{{
L\alpha}}^{\mathtt c}~{}^{({{10}}_{45})}\!{\bTau}^{c}_{L}~+~
{}^{({{10}}_{16})}\!{\bTau}^{c}_{L}~{}^{({{10}}_{45})}\!{\bf
{t}}_{{ L\alpha}}^{\mathtt c}
\right),\nonumber\\
\nonumber\\
K_{3}^{\alpha}&=&q\left[\left(-80\zeta_{\acute{a}\acute{b}}^{^{(10)(+)}}
+\frac{3}{5}\varrho_{\acute{a}\acute{b}}^{^{(126,\overline{126})(+)}}\right)
\left(\epsilon^{\alpha\beta\gamma}{\bf{D}}_{L\acute
{a}\beta}^{\mathsf c}{\bf{U}}_{L\acute {b}\gamma}^{\mathsf
c}\right)\right.\nonumber\\
&&\left.+2\left(40\zeta_{\acute{a}\acute{b}}^{^{(10)(+)}}
-\frac{1}{15}\varrho_{\acute{a}\acute{b}}^{^{(126,\overline{126})(+)}}\right)
\left({\bf E}_{L\acute {a}}{\bf{U}}^{\alpha}_{L\acute
{b}}+{\Nu}_{L\acute {a}}{\bf{D}}^{\alpha}_{L\acute
{b}}\right)\right]\nonumber\\
&&+\frac{1}{2\sqrt{2}}f^{(10)}_{33}\left(\epsilon^{\alpha\beta\gamma}~{}^{({{10}}_{16})}\!{\bf
{t}}_{{ L\beta}}^{\mathtt c}~{}^{({\overline{5}}_{10})}\!{\bf
{b}}_{{ L\gamma}}^{\mathtt c}~-~{}^{({{10}}_{16})}\!{\bf {t}}_{{
L}}^{\alpha}~{}^{({\overline{5}}_{10})}\!{\bTau}
_{L}~-~{}^{({{10}}_{16})}\!{\bf {b}}_{{
L}}^{\alpha}~{}^{({\overline{5}}_{10})}\!{\bNu} _{L\tau}\right),\nonumber\\
\nonumber\\
J_{4}^{\alpha}&=&-8p\lambda_{\acute{a},\acute{b}}^{^{(45)}}\left(\epsilon^{\alpha\beta\gamma}
{\bf{U}}_{L\acute {a}\beta}^{\mathsf c}{\bf{U}}_{L\acute
{b}\gamma}^{\mathsf c}
\right)-f^{(45)}_{33}\left(\epsilon^{\alpha\beta\gamma}~{}^{({{10}}_{16})}\!{\bf
{t}}_{{ L\beta}}^{\mathtt c}~{}^{({{10}}_{45})}\!{\bf {t}}_{{
L\gamma}}^{\mathtt c}\right),\nonumber\\
\nonumber\\
K_{4\alpha}&=&-\frac{16q}{15}\varrho_{\acute{a}\acute{b}}^{^{(126,\overline{126})(+)}}
\left({\bf{D}}_{L\acute {a}\alpha}^{\mathsf c}{\bf E}^{c
}_{L\acute {b}}\right)+\frac{1}{\sqrt
2}f^{(10)}_{33}\left({}^{({{10}}_{16})}\!{\bTau}^{c
}_{L}~{}^{({\overline{5}}_{10})}\!{\bf {b}}_{{ L\alpha}}^{\mathsf
c}\right).
\end{eqnarray}
 In the above ${\acute{a}, \acute{b}}$ etc stand for the first two generations while 
the third generation is explicitly factored out and the fields denoted by their 
familiar symbols 
${\bf b}, 
{\bf t}, {\bf\tau}$ for the bottom quark, top quark and for the $\tau$ lepton.
We note that the currents $J_4$ and $K_4$ are similar  to the
tilde currents $\tilde J$ and $\tilde K$ discussed in
Sec.~\ref{sec3}. Finally, integrating out the Higgs triplet fields
in Eq.(\ref{B-L}), we obtain the usual $RRRR$ and $LLLL$ operators.
These are exhibited in Appendix B.\\
 We now explore the conditions under  which proton decay is
suppressed. First all the terms which contain tau do  not
contribute  since  proton cannot decay into a final state with a
tau. 
 It is now easily checked that
all the remaining $LLLL$ and $RRRR$ terms do not contribute or cancel under the
constraints 
\beqn 
\lambda_{\acute{a}\acute{b},\acute{c}\acute{d}}^{^{(45)}}=
4\xi_{\acute{a}\acute{b}, \acute{c}\acute{d}}^{^{(10)(+)}},
~~\zeta_{\acute{a}\acute{b},\acute{c}\acute{d}}^{^{(10)(+)}} =0=
\lambda_{\acute{a}\acute{b},\acute{c}\acute{d}}^{^{(10)(+)}} =
\varrho_{\acute{a}\acute{b},\acute{c}\acute{d}}^{^{(126,\overline{126})(+)}}.
\label{cancel2} \eeqn
The cancellation condition of  Eq.(\ref{cancel2}) involves the
 quartic couplings and thus the Planck scale effects.
A cancellation to reduce the  proton decay  amplitude  by a  factor  of  ten
will lead to extending the proton life  time by a factor $10^2$ and
may  be sufficient to suppress proton decay  to the current level
of experiment for most models making them  phenomenologically  viable.
At the same time it leaves open the possibility that   proton decay may be
observed  in the next round of experiment\cite{p-future}.
  For the case when the $Q=-1/3(1/3)$
Higgsino exchange is much more suppressed  than the $Q=-4/3(4/3)$
Higgsino exchange, there will be only $RRRR$ type operators and
the dominant mode would be $ \bar\nu_{\tau} \bar{K}^+$.
%%%%%%%%%%%%%%%%%%%%%%%
\subsection{quark-lepton textures}
The analysis presented above  produces  the quark lepton
textures with the correct sizes.
The first two generation of masses arise  in the above  model from quartic
couplings  of matter fields  with the $144$ and $\overline{144}$  Higgs  fields\cite{bgns1},
while
the third generation masses  arise  from cubic couplings  involving  $144$ and
$\overline{144}$ of Higgs  and additional $10$ and $45$ plets of 
$SO(10)$
matter fields\cite{bgns2}.
Below we show how  the quartic couplings can generate  the desired sizes  for  the
up quark, down quark, lepton and neutirno masses for the first two generations.
Using the decomposition of $SO(10)$ couplings in terms of $SU(5)$ couplings  
the up-quark masses arise from the interactions
\beqn
10_M  10_M \frac{24_H}{M}5_H
\eeqn
where the fields above are all in $SU(5)$ representations.  Similarly
 the down quark and lepton  masses arise from the interaction
\beqn
10_M\bar 5_M \frac{24_H}{M}\bar 5_H
\eeqn
The $RR$, $LR$ and $LL$ neutrino masses arise from the following terms
\beqn
{\rm  RR}-\nu {\rm ~mass}:~~1_M1_M \frac{24_H}{M}24_H\nonumber\\
{\rm LR}-\nu {\rm ~mass}:~~\bar 5_M1_M \frac{24_H}{M} (5_H, 45_H)\nonumber\\
{\rm LL}-\nu {\rm ~mass}:~~\bar 5_M \bar 5_M\frac{5_H}{M} 5_H
\eeqn
It is now easily seen that  the right sizes for the quark-lepton masses  for the first two
generations can appear
after $24_H$, $5_H+\bar 5_H$, $45_H+\overline{45}_H$ develop
vacuum expectation values. Additionally for the third generation one has cubic couplings
which can generate relatively large masses typical of third generation.
The full analysis is rather  involved and is outside the
scope of this paper.

\section{CONCLUSION\label{sec6}}
In this paper we have discussed  the mechanism where contributions
from different operators that contribute to the baryon and lepton
number  violating dimension 5 operators tend to cancel producing
an enhancement for the proton decay lifetime in supersymmetric
unified theories. The cancellation mechanism works when there are
more than one pair of Higgs triplets generating baryon and lepton
number  violating interactions,  with their Yukawa coupling having
similar  generational symmetry. We have discussed in this paper
three specific examples,
 two for the $SU(5)$ case and the other for the  $SO(10)$ case.
For the $SU(5)$ case we first considered  a model with a Higgs sector
consisting of $5_H+\bar 5_H$, $24_H$, and $45_H+\overline{45}_H$
plets of Higgs. Here  the $24_H$ plet breaks the GUT symmetry down
to $SU(3)_C\times SU(2)_L\times U(1)_Y$, and the Higgs doublets
from the $5_H+\bar 5_H$ enter in the electroweak symmetry
breaking, while the Higgs triplet fields from $5_H+\bar 5_H$ and
from $45_H+\overline{45}_H$ generate baryon and lepton number
violating interactions. It is then shown that the  baryon and
lepton number  violating contributions  arising from the exchange
of Higgs triplets from the $45_H+\overline{45}_H$ can cancel  the
baryon and lepton number  contributions arising from the  Higgs
triplet exchange from the $5_H + \bar 5_H$ when the  generational
dependence of the Yukawa couplings of the
$5_H + \bar 5_H$ and of  $45_H + \overline{45}_H$ are  similar.
Next we considered  an $SU(5)$ example with only $5_H+\bar 5_H$
and
$24_H$ of Higgs but including  Planck scale contributions . Here it is
seen that one can produce the appropriate quark-lepton textures and
a complete suppression of  B\&L violating dimension five operators.
\\

For the $SO(10)$ case, we consider  a  recently proposed model
where a  one step breaking of $SO(10)$ to $SU(3)_C\times
U(1)_{em}$ can occur with a $144_H+\overline{144}_H$  of Higgs.
Here the decomposition of $144_H+\overline{144}_H$
 contains automatically $45_H+\overline{45}_H$  of Higgs in addition to $5_H+\bar 5_H$ of Higgs.
The   interactions of $144_H+\overline{144}_H$  with matter are at
least  quartic, but normal Yukawa  type coupling arise after
spontaneous breaking when $144_H+\overline{144}_H$  develop VEVs.
In addition   large 3rd generation masses can arise with cubic
interactions  when  $10 + 45$ of matter is included. The analysis
including all these interactions was carried  out and baryon and
lepton number violating dimension 5 operators were computed. It is
then found that simple constraints suppress all LLLL  and RRRR
baryon and lepton number  violating dimension five operators.
While  we have illustrated the mechanism for three models, it is
likely applicable to a larger class  in which the baryon and
lepton number  violating operators arise from more than one
source.
 The cancellation mechanism can allow for a complete or
partial  suppression  of proton decay, allowing for suppression
consistent with the current experimental limits while allowing for
the possibility that proton decay may become visible in the next
round of nucleon stability experiments\cite{p-future}.
 Finally, we
have not addressed in this work issues related to mass spectra for
the heavy fields, gauge coupling unification,  and a detailed numerical
fit to quark-lepton-neutrino mass textures.  A detailed analysis of
these topics is outside the scope of this paper. These topics are
worthy of further investigations. \\

Finally we note that recently there has been much further work on the
interface of GUTs and strings  (see, e.g., Ref.\cite{Lebedev:2007hv}
and the references  therein)
which make  progress towards the generation of realistic particle physics
models.  While some models are  free of  B\&L violating dimension
five operators, others are not\cite{Lebedev:2007hv},
and the cancellation mechanism may play a role in making
such models viable. More specifically, a class of string models which
would otherwise be  eliminated by the experimental constraint on
B\&L violating dimension
five  operators  could become phenomenology admissible
using the cancellation mechanism proposed here.
Further, as noted  in Sec.(II)  the quantum gravity corrections which were introduced in 
Eqs.(4) and (5) could also be utilized for the suppression of dimension five proton
decay. However, a detailed analysis  of this phenomenon is outside  the scope
of the this paper.

%We further note,
%that a detailed analysis of these  issues is not necessary for the
%imposition of the cancellation mechanism. Nonetheless,  the
%computation of the mass spectra of heavy fields, for example, in
%$144_H+\overline{144}_H$, analysis of gauge coupling unification
%with inclusion of these thresholds, and an analysis of
%quark-lepton-neutrino textures are all topics worthy of further
%investigation.\\

\begin{acknowledgments}
Correspondence with K.S. Babu and Ilia Gogoladze in early stage
of the work is acknowledged. This work is supported in part by NSF
grant  PHY-0456568.
\end{acknowledgments}
\section{Appendices}
\subsection{Matter -Higgs quartic  couplings}
In this appendix we  give  further details of the matter-higgs quartic couplings discussed in
Sec.(VC).  Below we exhibit the results for $W_4^{(i)}$ (i=1-5) that appear in Sec.(VC).
Our analysis  gives
 \begin{widetext}
\begin{eqnarray}
W^{(1)}_4= {\bf M}_{\acute{a}}{\bf M}_{\acute{b}}\left\{\left[
-\lambda_{\acute{a}\acute{c},\acute{b}\acute{d}}^{^{(45)}}
+\lambda_{\acute{a}\acute{c},\acute{b}\acute{d}}^{^{(54)}}
\right]{{\bf P}}_{\acute{c}j}^i{{\bf P}}_{\acute{d}i}^j
+\frac{4}{\sqrt 5}\left[
\frac{4}{15}\varrho_{\acute{a}\acute{b},\acute{c}\acute{d}}^{^{(126,\overline{126})(+)}}+
\lambda_{\acute{a}\acute{c},\acute{b}\acute{d}}^{^{(10)}}
\right]{{\bf Q}}_{\acute{c}}^i{{\bf Q}}_{\acute{d}i}\right\},
\end{eqnarray}
\begin{eqnarray}
W^{(2)}_4= &&{\bf M}_{\acute{a}i}{\bf
M}_{\acute{b}}\left\{\frac{1}{\sqrt 5}\left[
-3\lambda_{\acute{a}\acute{c},\acute{b}\acute{d}}^{^{(45)}}
+\lambda_{\acute{a}\acute{c},\acute{b}\acute{d}}^{^{(54)}}
+8\xi_{\acute{a}\acute{b},\acute{c}\acute{d}}^{^{(10)(+)}}+\frac{8}{3}\xi_{\acute{a}\acute{b},\acute{c}\acute{d}}^{^{(120)(-)}}\right]{{\bf
P}}_{\acute{c}}^j{{\bf P}}_{\acute{d}j}^i\right.\nonumber\\
&&\left.+2\left[
-\lambda_{\acute{a}\acute{c},\acute{b}\acute{d}}^{^{(45)}}
-\lambda_{\acute{a}\acute{c},\acute{b}\acute{d}}^{^{(54)}}
+8\xi_{\acute{a}\acute{b},\acute{c}\acute{d}}^{^{(10)(+)}}+\frac{8}{3}\xi_{\acute{a}\acute{b},\acute{c}\acute{d}}^{^{(120)(-)}}\right]{{\bf
P}}_{\acute{c}k}^{ij}{{\bf P}}_{\acute{d}j}^k\right.\nonumber\\
&&\left.+2\left[\frac{16}{5}\varrho_{\acute{a}\acute{b},\acute{c}\acute{d}}^{^{(126,\overline{126})(+)}}
+\lambda_{\acute{a}\acute{d},\acute{b}\acute{c}}^{^{(10)}}
-8\zeta_{\acute{a}\acute{b},\acute{c}\acute{d}}^{^{(10)(+)}}
+\frac{8}{3}\zeta_{\acute{a}\acute{b},\acute{c}\acute{d}}^{^{(120)(-)}}\right]{{\bf
Q}}_{\acute{c}}^{j}{{\bf Q}}_{\acute{d}j}^i\right \},
\end{eqnarray}
\begin{eqnarray}
W^{(3)}_4= {\bf M}_{\acute{a}i}{\bf M}_{\acute{b}j}{{\bf
P}}_{\acute{c}}^{i}{{\bf P}}_{\acute{d}}^j\left[
\frac{1}{4}\left(\lambda_{\acute{a}\acute{c},\acute{b}\acute{d}}^{^{(45)}}
-\frac{201}{25}\lambda_{\acute{a}\acute{c},\acute{b}\acute{d}}^{^{(54)}}
\right)-\frac{1}{4}\left(5\lambda_{\acute{a}\acute{d},\acute{b}\acute{c}}^{^{(45)}}
-\frac{9}{5}\lambda_{\acute{a}\acute{d},\acute{b}\acute{c}}^{^{(54)}}
\right)-\frac{32}{15}\xi_{\acute{a}\acute{b},\acute{c}\acute{d}}^{^{(120)(-)}}
\right],
\end{eqnarray}
\begin{eqnarray}\label{candidates down}
W^{(4)}_4= &&{\bf M}_{\acute{a}}^{ij}{\bf
M}_{\acute{b}j}\left\{2\left[
-\lambda_{\acute{a}\acute{c},\acute{b}\acute{d}}^{^{(45)}}
-\lambda_{\acute{a}\acute{c},\acute{b}\acute{d}}^{^{(54)}}
+8\xi_{\acute{a}\acute{b},\acute{c}\acute{d}}^{^{(10)(+)}}+\frac{8}{3}\xi_{\acute{a}\acute{b},\acute{c}\acute{d}}^{^{(120)(-)}}\right]{{\bf
P}}_{\acute{c}i}^k{{\bf P}}_{\acute{d}k}\right.\nonumber\\
&&\left.+\frac{1}{\sqrt
5}\left[\frac{1}{15}\varrho_{\acute{a}\acute{b},\acute{c}\acute{d}}^{^{(126,\overline{126})(+)}}
+8\zeta_{\acute{a}\acute{b},\acute{c}\acute{d}}^{^{(10)(+)}}
+\frac{8}{3}\zeta_{\acute{a}\acute{b},\acute{c}\acute{d}}^{^{(120)(-)}}\right]{{\bf
Q}}_{\acute{c}k}{{\bf Q}}_{\acute{d}j}^k \right.\nonumber\\
&&\left.+2\left[\frac{1}{15}\varrho_{\acute{a}\acute{b},\acute{c}\acute{d}}^{^{(126,\overline{126})(+)}}
+8\zeta_{\acute{a}\acute{b},\acute{c}\acute{d}}^{^{(10)(+)}}
-\frac{8}{3}\zeta_{\acute{a}\acute{b},\acute{c}\acute{d}}^{^{(120)(-)}}\right]{{\bf
Q}}_{\acute{c}ik}^l{{\bf Q}}_{\acute{d}l}^k\right\}\nonumber\\
&&+{\bf M}_{\acute{a}}^{ij}{\bf
M}_{\acute{b}k}\left\{\frac{1}{\sqrt
5}\left[\frac{4}{15}\varrho_{\acute{a}\acute{b},\acute{c}\acute{d}}^{^{(126,\overline{126})(+)}}
-\lambda_{\acute{a}\acute{c},\acute{b}\acute{d}}^{^{(10)}}
+\frac{16}{3}\zeta_{\acute{a}\acute{b},\acute{c}\acute{d}}^{^{(120)(-)}}\right]{\bf
Q}_{\acute{c}j}{{\bf Q}}_{\acute{d}i}^k \right.\nonumber\\
&&\left.+\frac{4}{
3}\left[-\frac{1}{5}\varrho_{\acute{a}\acute{b},\acute{c}\acute{d}}^{^{(126,\overline{126})(+)}}
+4\zeta_{\acute{a}\acute{b},\acute{c}\acute{d}}^{^{(120)(-)}}\right]{\bf
Q}_{\acute{c}ij}^l{{\bf Q}}_{\acute{d}l}^k \right.\nonumber\\
&&\left.+2\left[-\lambda_{\acute{a}\acute{c},\acute{b}\acute{d}}^{^{(45)}}
+\lambda_{\acute{a}\acute{c},\acute{b}\acute{d}}^{^{(54)}}\right]
{\bf P}_{\acute{c}j}^k{{\bf P}}_{\acute{d}i}\right\},
\end{eqnarray}
\begin{eqnarray}\label{candidates up}
W^{(5)}_4= &&\epsilon_{ijklm}{\bf M}_{\acute{a}}^{ij}{\bf
M}_{\acute{b}}^{kl}\left\{2\left[
\frac{2}{15}\varrho_{\acute{a}\acute{b},\acute{c}\acute{d}}^{^{(126,\overline{126})(+)}}
+\zeta_{\acute{a}\acute{b},\acute{c}\acute{d}}^{^{(10)(+)}}+\frac{2}{3}\zeta_{\acute{a}\acute{b},\acute{c}\acute{d}}^{^{(120)(-)}}\right]{{\bf
Q}}_{\acute{c}}^n{{\bf Q}}_{\acute{d}n}^m\right.\nonumber\\
&&\left.+2\left[\xi_{\acute{a}\acute{b},\acute{c}\acute{d}}^{^{(10)(+)}}\right]{\bf
P}_{\acute{c}n}^p{{\bf P}}_{\acute{d}p}^{nm} -\frac{1}{\sqrt
5}\left[\xi_{\acute{a}\acute{b},\acute{c}\acute{d}}^{^{(10)(+)}}\right]{\bf
P}_{\acute{c}n}^m{{\bf P}}_{\acute{d}}^{n}\right\}\nonumber\\
&&+\epsilon_{ijklm}{\bf M}_{\acute{a}}^{in}{\bf
M}_{\acute{b}}^{jk}\left\{\left[-\frac{1}{2}\lambda_{\acute{a}\acute{c},\acute{b}\acute{d}}^{^{(45)}}
-\frac{1}{2}\lambda_{\acute{a}\acute{c},\acute{b}\acute{d}}^{^{(54)}}
+\frac{4}{3}\xi_{\acute{a}\acute{b},\acute{c}\acute{d}}^{^{(120)(-)}}\right]{\bf
P}_{\acute{c}n}^p{{\bf P}}_{\acute{d}p}^{lm}\right.\nonumber\\
&&\left.+\frac{1}{\sqrt
5}\left[-\lambda_{\acute{a}\acute{c},\acute{b}\acute{d}}^{^{(45)}}+\frac{4}{3}
\xi_{\acute{a}\acute{b},\acute{c}\acute{d}}^{^{(120)(-)}} \right]
{\bf P}_{\acute{c}n}^l{{\bf P}}_{\acute{d}}^{m}\right\}\nonumber\\
&&+\frac{1}{2}\epsilon_{ijklm}{\bf M}_{\acute{a}}^{np}{\bf
M}_{\acute{b}}^{ij}\left[\lambda_{\acute{a}\acute{c},\acute{b}\acute{d}}^{^{(45)}}
-\lambda_{\acute{a}\acute{c},\acute{b}\acute{d}}^{^{(54)}}\right]{\bf
P}_{\acute{c}p}^k{{\bf P}}_{\acute{d}n}^{lm}.
\end{eqnarray}
\end{widetext}

%%%%
\subsection{Analysis of LLLL and RRRR dimension five operators}
Below  we exhibit the result of the analysis of LLL and RRRR dimension five
operators.  These dimension five operators  are gotten by integration over
all the Higgs triplet fields.
\begin{eqnarray}\label{basicdim5}
{\mathsf W}_{B\&L }^{dim-5}&=&\sum_{g=I}^{III}\left({\mathsf
W}_{{\mathtt R}}^{(g)}+{\mathsf W}_{{\mathtt L}}^{(g)}\right).
\end{eqnarray}
The index $g$ above denotes whether the particular operator
connects one ($I$), two ($II$) or three ($III$) generations of
fermions. The operators in Eq.(\ref{basicdim5}) are defined
through
\begin{eqnarray}
{\mathsf W}_{{\mathtt R}}^{(I)}&=&
{\mathtt{R}}^{(I)}~~~\epsilon^{\alpha\beta\gamma}~{}^{({\overline{5}}_{16})}\!{\bf
{b}}_{{ L\alpha}}^{\mathtt c}~{}^{(10_{16})}\!{\bf {t}}_{{
L\beta}}^{\mathtt c}~{}^{(10_{16})}\!{\bf {t}}_{{
L\gamma}}^{\mathtt c}~{}^{(10_{16})}\!{\bTau}^{c }_{L},\\
 \nonumber\\
{\mathsf W}_{{\mathtt R}}^{(II)}&=& {\mathtt {R}}_{1~\acute
{a}\acute {b},\acute {c}\acute
{d}}^{(II)}~~~\epsilon^{\alpha\beta\gamma}~{\bf{D}}_{L\acute
{a}\alpha}^{\mathtt c}~{\bf{U}}_{L\acute {b}\beta}^{\mathtt
c}~{\bf{U}}_{L\acute {c}\gamma}^{\mathtt c}~{\bf E}^{c
}_{L\acute {d}} \nonumber\\
&+&{\mathtt {R}}_{2~\acute {a}\acute {b},\acute {c}\acute
{d}}^{(II)}~~~\epsilon^{\alpha\beta\gamma}~{\bf{D}}_{L\acute
{a}\alpha}^{\mathtt c}~{\bf E}^{c}_{L\acute
{b}}~{\bf{U}}_{L\acute {c}\beta}^{\mathtt c}~{\bf{U}}_{L\acute
{d}\gamma}^{\mathtt
c},\\
 \nonumber\\
 {\mathsf W}_{{\mathtt R}}^{(III)}&=&
{\mathtt {R}}_{1~\acute {a}\acute
{b}}^{(III)}~~~\epsilon^{\alpha\beta\gamma}~
~{}^{({\overline{5}}_{16})}\!{\bf {b}}_{{ L\alpha}}^{\mathtt c}
~{}^{(10_{16})}\!{\bf {t}}_{{ L\beta}}^{\mathsf
c}~{\bf{U}}_{L\acute {a}\gamma}^{\mathsf c}~{\bf E}^{c
}_{L\acute {b}}\nonumber\\
&+&{\mathtt {R}}_{2~\acute {a}\acute
{b}}^{(III)}~~~\epsilon^{\alpha\beta\gamma}~ {\bf{D}}_{L\acute
{a}\alpha}^{\mathsf c}~{\bf{U}}_{L\acute {b}\beta}^{\mathsf
c}~{}^{(10_{16})}\!{\bf {t}}_{{ L\gamma}}^{\mathtt
c}~{}^{(10_{16})}\!{\bTau}^{c}_{L}
\nonumber\\
 &+&
{\mathtt {R}}_{3~\acute {a}\acute
{b}}^{(III)}~~~\epsilon^{\alpha\beta\gamma}~
~{}^{({\overline{5}}_{16})}\!{\bf {b}}_{{ L\alpha}}^{\mathtt
c}~{}^{({{10}}_{16})}\!{\bTau}^{c}_{L}~{\bf{U}}_{L\acute
{a}\beta}^{\mathtt c}~{\bf{U}}_{L\acute {b}\gamma}^{\mathsf
c}\nonumber\\
&+&{\mathtt{R}}_{4~\acute {a}\acute
{b}}^{(III)}~~~\epsilon^{\alpha\beta\gamma}~{\bf{D}}_{L\acute
{a}\alpha}^{\mathtt c}~{\bf E}^{c}_{L\acute
{b}}~{}^{({{10}}_{16})}{\bf {t}}_{{ L\beta}}^{\mathtt
c}~{}^{(10_{16})}\!{\bf {t}}_{{ L\gamma}}^{\mathtt c},\\
\nonumber\\
\nonumber\\
 {\mathsf W}_{{\mathtt L}}^{(I)}&=&
{\mathtt{L}}^{(I)}_1~~~\epsilon_{\alpha\beta\gamma}~{}^{({\overline{5}}_{16})}\!{\bTau}
_{L}~{}^{({{10}}_{16})}\!{\bf {t}}_{{
L}}^{\alpha}~{}^{({{10}}_{16})}\!{\bf {t}}_{{
L}}^{\beta}~{}^{(10_{16})}\!{\bf {b}}_{{ L}}^{\gamma}
\nonumber\\
&+&{\mathtt{L}}^{(I)}_2~~~\epsilon_{\alpha\beta\gamma}~
^{({\overline{5}}_{16})}\!{\bNu}
_{L\tau}~{}^{({{10}}_{16})}\!{\bf {b}}_{{ L}}^{\alpha}
~{}^{({{10}}_{16})}\!{\bf {t}}_{{
L}}^{\beta}~{}^{(10_{16})}\!{\bf {b}}_{{ L}}^{\gamma},\\
\nonumber\\
{\mathsf W}_{{\mathtt L}}^{(II)}&=& {\mathtt{L}}^{(II)}_{\acute
{a}\acute {b},\acute {c}\acute
{d}}~~~\left[\epsilon_{\alpha\beta\gamma}~ {\bf E}_{L\acute
{a}}~{\bf{U}}_{L\acute {b}}^{\alpha}~{\bf{U}}_{L\acute
{c}}^{\beta}~{\bf{D}}_{L\acute
{d}}^{\gamma}~+~\epsilon_{\alpha\beta\gamma}~{\Nu}_{L\acute
{a}}~{\bf{D}}_{L\acute {b}}^{\alpha}~{\bf{U}}_{L\acute
{c}}^{\beta}~{\bf{D}}_{L\acute
{d}}^{\gamma} \right],\\
\nonumber\\
{\mathsf W}_{{\mathtt L}}^{(III)}&=&
{\mathtt{L}}^{(III)}_{1~\acute {a}\acute
{b}}~~~\left[\epsilon_{\alpha\beta\gamma}~{\bf E}_{L\acute
{a}}~{\bf{U}}_{L\acute {b}\alpha}~{}^{({{10}}_{16})}\!{\bf
{t}}_{{ L}}^{\beta}{}~^{(10_{16})}\!{\bf {b}}_{{ L}}^{\gamma}
~+~\epsilon_{\alpha\beta\gamma}~{\Nu}_{L\acute
{a}}~{\bf{D}}_{L\acute {b}}^{\alpha}~{}^{({{10}}_{16})}\!{\bf
{t}}_{{ L}}^{\beta}{}~^{(10_{16})}\!{\bf {b}}_{{
L}}^{\gamma}\right]\nonumber\\
&+& {\mathtt{L}}^{(III)}_{2~\acute {a}\acute
{b}}~~~\epsilon_{\alpha\beta\gamma}~{}^{({\overline{5}}_{16})}\!{\bTau}
_{L}~{}^{(10_{16})}\!{\bf {t}}_{{ L}}^{\alpha}~{\bf{U}}_{L\acute
{a}}^{\beta}~{\bf{D}}_{L\acute {b}}^{\gamma}\nonumber\\
&+& {\mathtt{L}}^{(III)}_{2~\acute {a}\acute
{b}}~~~\epsilon_{\alpha\beta\gamma}~{}^{({\overline{5}}_{16})}\!{\bNu}
_{L\tau}~{}^{(10_{16})}\!{\bf {b}}_{{
L}}^{\alpha}~{\bf{U}}_{L\acute {a}}^{\beta}~{\bf{D}}_{L\acute
{b}}^{\gamma}.
\end{eqnarray}
The coefficients ${\mathtt{L}}$ and ${\mathtt{R}}$ are defined in
Tables  1 and 2.  In computing them we  have  limited  ourselves  to one
generation of $144+\overline{144}$ plet of Higgs. Thus the couplings  with
120 plet mediation which are  anti-symmetric in the generation indices
vanish.
\\

 {\small{
\begin{center} \begin{tabular}{||c||l|}
\multicolumn{2}{c}{Table 1 : Definition of parameters in Table
2}\cr

\hline
    ${\mathtt A}$ & $\frac{\cos^2\vartheta_{\mathsf T}}{M_{{\mathsf
T}_2}}+\frac{\sin^2\vartheta_{\mathsf T}}{M_{{\mathsf T}_3}}$\\
\hline

    ${\mathtt B}$ & $\frac{\sin^2\vartheta_{\mathsf T}}{M_{{\mathsf
T}_2}}+\frac{\cos^2\vartheta_{\mathsf T}}{M_{{\mathsf T}_3}}$\\
\hline

    ${\mathtt C}$ & $\left(\frac{1}{M_{{\mathsf
T}_2}}-\frac{1}{M_{{\mathsf T}_3}}\right)\cos\vartheta_{\mathsf T}\sin\vartheta_{\mathsf T}$\\
\hline
    ${\mathtt X}_{1\acute {a}\acute {b}}$ & $8p\left(4\xi_{\acute{a}\acute{b}}^{^{(10)(+)}}
-\lambda_{\acute{a},\acute{b}}^{^{(45)}}\right)$\\

\hline
    ${\mathtt X}_{2\acute {a}\acute {b}}$ & $2p\left(-16\xi_{\acute{a}\acute{b}}^{^{(10)(+)}}
-\lambda_{\acute{a},\acute{b}}^{^{(45)}}+5\lambda_{\acute{a},\acute{b}}^{^{(54)}}\right)$\\

\hline
    ${\mathtt X}_{3\acute {a}\acute {b}}$ & $20p\left(-4\xi_{\acute{a}\acute{b}}^{^{(10)(+)}}
+\lambda_{\acute{a},\acute{b}}^{^{(54)}}\right)$\\

\hline
    ${\mathtt Y}_{1\acute {a}\acute {b}}$ & $32q\left(\zeta_{\acute{a}\acute{b}}^{^{(10)(+)}}+\frac{2}{15}
\varrho_{\acute{a}\acute{b}}^{^{(126,\overline{126})(+)}}\right)$\\

\hline
    ${\mathtt Y}_{2\acute {a}\acute {b}}$ & $\frac{q}{\sqrt
5}\left(16\zeta_{\acute{a}\acute{b}}^{^{(10)(+)}}
+2\lambda_{\acute{a},\acute{b}}^{^{(10)}}
-\frac{3}{5}\varrho_{\acute{a}\acute{b}}^{^{(126,\overline{126})(+)}}\right)$\\

\hline
    ${\mathtt Y}_{3\acute {a}\acute {b}}$ & $\frac{q}{\sqrt
5}\left(-16\zeta_{\acute{a}\acute{b}}^{^{(10)(+)}}
+3\lambda_{\acute{a},\acute{b}}^{^{(10)}}
-\frac{14}{15}\varrho_{\acute{a}\acute{b}}^{^{(126,\overline{126})(+)}}\right)$\\

\hline
    ${\mathtt Y}_{4\acute {a}\acute {b}}$ & $q\left(-80\zeta_{\acute{a}\acute{b}}^{^{(10)(+)}}
+\frac{3}{5}\varrho_{\acute{a}\acute{b}}^{^{(126,\overline{126})(+)}}\right)$\\

\hline
    ${\mathtt Y}_{5\acute {a}\acute {b}}$ & $2q\left(40\zeta_{\acute{a}\acute{b}}^{^{(10)(+)}}
-\frac{1}{5}\varrho_{\acute{a}\acute{b}}^{^{(126,\overline{126})(+)}}\right)$\\
\hline
\end{tabular}
\end{center}
}}

{\small{
\begin{center} \begin{tabular}{||c||l|}
\multicolumn{2}{c}{Table 2 : Coefficients of $LLLL$ and $RRRR$
baryon and lepton number  violating  dimension five operators}\cr
\hline

 ${\mathtt{R}}^{(I)}$ &
$f^{(10)}_{33}f^{(45)}_{33}\sin\theta_{u\textnormal{b}}\cos\theta_{u\textnormal
{t}}\left[\left(-\frac{1}{5}{\mathtt A}-{\mathtt B}+\frac{2}{\sqrt
5}{\mathtt C}\right)\sin\theta_{u\mTau}\cos\theta_{u\textnormal
{t}} +\left(-\frac{1}{5}{\mathtt A}+ {\mathtt B}+\frac{1}{\sqrt
{2}}\frac{1}{M_{{\mathsf
T}_4}}\right)\cos\theta_{u\mTau}\sin\theta_{u\textnormal{t}}\right]
$ \\

\hline

${\mathtt{R}}_{1~\acute {a}\acute {b},\acute {c}\acute
{d}}^{(II)}$&$-\frac{1}{M_{{\mathsf T}_1}}{\mathtt X}_{1\acute
{a}\acute {b}}{\mathtt Y}_{1\acute {c}\acute
{d}}+\left[\left(\frac{1}{2\sqrt{5}}{\mathtt
A}-\frac{5}{2}{\mathtt C}\right){\mathtt Y}_{2\acute {a}\acute
{b}}+\left(\frac{1}{2\sqrt{5}}{\mathtt C}-\frac{5}{2}{\mathtt
B}\right){\mathtt Y}_{4\acute {a}\acute {b}}\right]{\mathtt
X}_{1\acute
{c}\acute {d}}$\\

\hline

 ${\mathtt{R}}_{2~\acute {a}\acute {b},\acute {c}\acute
{d}}^{(II)}$&$-\frac{128}{15}\frac{pq}{M_{{\mathsf T}_4}}
\varrho_{\acute{a}\acute{b}}^{^{(126,\overline{126})(+)}}
\lambda_{\acute{a},\acute{b}}^{^{(45)}}$ \\

\hline

${\mathtt{R}}_{1~\acute {a}\acute
{b}}^{(III)}$&$2\sqrt{2}\frac{1}{M_{{\mathsf
T}_1}}f^{(45)}_{33}\cos\theta_{u\textnormal
{b}}\sin\theta_{u\textnormal{t}}~{\mathtt Y}_{1\acute {a}\acute
{b}}-f^{(10)}_{33}\left(\frac{1}{20\sqrt{2}}{\mathtt
A}+\frac{5}{4\sqrt{2}}{\mathtt B}-\frac{3}{2\sqrt{10}}{\mathtt
C}\right)
\sin\theta_{u\textnormal{b}}\cos\theta_{u\textnormal{t}}~{\mathtt
X}_{1\acute
{a}\acute {b}}$ \\

\hline

${\mathtt{R}}_{2~\acute {a}\acute {b}}^{(III)}$&
$f^{(45)}_{33}\left\{\left[\left(2\sqrt{\frac{2}{{5}}}{\mathtt
A}-2\sqrt{2}{\mathtt C}\right){\mathtt Y}_{2\acute {a}\acute
{b}}+\left(2\sqrt{\frac{2}{{5}}}{\mathtt C}-2\sqrt{2}{\mathtt
B}\right){\mathtt Y}_{4\acute {a}\acute
{b}}\right]\cos\theta_{u\textnormal
{t}}\sin\theta_{u\mTau}\right.$\\
& $\left. +\left[\left(2\sqrt{\frac{2}{{5}}}{\mathtt
A}+2\sqrt{2}{\mathtt C}\right){\mathtt Y}_{2\acute {a}\acute
{b}}+\left(2\sqrt{\frac{2}{{5}}}{\mathtt C}+2\sqrt{2}{\mathtt
B}\right){\mathtt Y}_{4\acute {a}\acute
{b}}\right]\sin\theta_{u\textnormal {t}}\cos\theta_{u\mTau}
\right\}
$\\

\hline

${\mathtt{R}}_{3~\acute {a}\acute
{b}}^{(III)}$&$-4\sqrt{2}\frac{p}{M_{{\mathsf T}_4}}
f^{(10)}_{33}\sin\theta_{u\textnormal
{b}}\cos\theta_{u\mTau}~\lambda_{\acute{a},\acute{b}}^{^{(45)}}$\\

\hline

${\mathtt{R}}_{4~\acute {a}\acute {b}}^{(III)}$&
$\frac{16}{15}\frac{q}{M_{{\mathsf
T}_4}}f^{(45)}_{33}\cos\theta_{u\textnormal
{b}}\sin\theta_{u\mTau}~\varrho_{\acute{a}\acute{b}}^{^{(126,\overline{126})(+)}}$\\

\hline

${\mathtt{L}}^{(I)}_1$ &
$-f^{(10)}_{33}f^{(45)}_{33}\left(\frac{1}{5}{\mathtt
A}+\frac{1}{\sqrt 5}{\mathtt C}\right)\sin(\theta_{v\textnormal
{b}}+\theta_{v\textnormal
{t}})~\sin\theta_{v\mTau}\cos\theta_{v\textnormal
{t}}$\\

\hline

${\mathtt{L}}^{(I)}_2$ &
$-f^{(10)}_{33}f^{(45)}_{33}\left(\frac{1}{5}{\mathtt
A}+\frac{1}{\sqrt 5}{\mathtt C}\right)\sin(\theta_{v\textnormal
{b}}+\theta_{v\textnormal
{t}})~\sin\theta_{{v}{\mNu_{\tau}}}\cos\theta_{v\textnormal
{b}}$\\

\hline

${\mathtt{L}}_{\acute {a}\acute {b},\acute {c}\acute {d}}^{(II)}$&
$\frac{1}{M_{{\mathsf T}_1}}{\mathtt X}_{2\acute {a}\acute
{b}}{\mathtt Y}_{1\acute {c}\acute {d}}-Y_{3\acute {a}\acute
{b}}\left({\frac{1}{{2\sqrt{5}}}}{\mathtt A} {\mathtt X}_{1\acute
{c}\acute {d}}+{\mathtt C} {\mathtt X}_{3\acute {c}\acute
{d}}\right)-{\mathtt Y}_{5\acute {a}\acute
{b}}\left({\frac{1}{{2\sqrt{5}}}}{\mathtt C} {\mathtt X}_{1\acute
{c}\acute
{d}}+{\mathtt B} {\mathtt X}_{3\acute {c}\acute {d}}\right)$\\
 \hline

 ${\mathtt{L}}_{1~\acute {a}\acute {b}}^{(III)}$ & $2\sqrt{\frac{2}{5}}f^{(45)}_{33}
 \left({\mathtt A} {\mathtt Y}_{3\acute {a}\acute
{b}}+{\mathtt C}{\mathtt Y}_{5\acute {a}\acute
{b}}\right)\sin(\theta_{v\textnormal {b}}+\theta_{v\textnormal
{t}})$\\
\hline

${\mathtt{L}}_{2~\acute {a}\acute {b}}^{(III)}$&
$2\sqrt{2}\frac{1}{M_{{\mathsf
T}_1}}f^{(45)}_{33}\cos\theta_{v\mTau}\sin\theta_{v\textnormal
{t}}~{\mathtt Y}_{1\acute {a}\acute
{b}}-f^{(10)}_{33}\left[\left({\frac{1}{{20\sqrt{2}}}}{\mathtt
A}+{\frac{1}{{4\sqrt{10}}}}{\mathtt C}\right){\mathtt X}_{1\acute
{a}\acute {b}}+\left({\frac{1}{{2\sqrt{2}}}}{\mathtt
B}+{\frac{1}{{2\sqrt{10}}}}{\mathtt C}\right){\mathtt X}_{3\acute
{a}\acute {b}}\right]\sin\theta_{v\mTau}\cos\theta_{v\textnormal
{t}}$\\
\hline

${\mathtt{L}}_{3~\acute {a}\acute {b}}^{(III)}$&
$2\sqrt{2}\frac{1}{M_{{\mathsf
T}_1}}f^{(45)}_{33}\cos\theta_{{v}{\mNu_{\tau}}}
\sin\theta_{v\textnormal {b}}~{\mathtt Y}_{1\acute {a}\acute
{b}}-f^{(10)}_{33}\left[\left({\frac{1}{{20\sqrt{2}}}}{\mathtt
A}+{\frac{1}{{4\sqrt{10}}}}C\right){\mathtt X}_{1\acute {a}\acute
{b}}+\left({\frac{1}{{2\sqrt{2}}}}{\mathtt
B}+{\frac{1}{{2\sqrt{10}}}}{\mathtt C}\right){\mathtt X}_{3\acute
{a}\acute {b}}\right]\sin\theta_{{v}{\mNu_{\tau}}}
\cos\theta_{v\textnormal
{b}}$\\

\hline
\end{tabular}
\end{center}
}}
%\end{widetext}
%\input{bibl_sort.tex}


\begin{thebibliography}{999}

\bibitem{Pati:1974yy}
  J.~C.~Pati and A.~Salam,
  %``Lepton Number As The Fourth Color,''
  Phys.\ Rev.\  D {\bf 10}, 275 (1974)
  [Erratum-ibid.\  D {\bf 11}, 703 (1975)].
  %%CITATION = PHRVA,D10,275;%%

\bibitem{Georgi:1974sy}
  H.~Georgi and S.~L.~Glashow,
  %``Unity Of All Elementary Particle Forces,''
  Phys.\ Rev.\ Lett.\  {\bf 32}, 438 (1974).
  %%CITATION = PRLTA,32,438;%%

\bibitem{georgi}
H. Georgi, in Particles and Fields (edited by C.E. Carlson),
A.I.P., 1975; H. Fritzch and P. Minkowski, Ann. Phys. {\bf
93}(1975)193.

\bibitem{Nath:2006ut}
  P.~Nath and P.~F.~Perez,
  %``Proton stability in grand unified theories, in strings, and in branes,''
  Phys.\ Rept.\  {\bf 441}, 191 (2007)
  [arXiv:hep-ph/0601023].
  %%CITATION = PRPLC,441,191;%%

\bibitem{Dimopoulos:1981zb}
  S.~Dimopoulos and H.~Georgi,
  %``Softly Broken Supersymmetry And SU(5),''
  Nucl.\ Phys.\  B {\bf 193}, 150 (1981).
  %%CITATION = NUPHA,B193,150;%

\bibitem{Chamseddine:1982jx}
  A.~H.~Chamseddine, R.~Arnowitt and P.~Nath,
  %``Locally Supersymmetric Grand Unification,''
  Phys.\ Rev.\ Lett.\  {\bf 49}, 970 (1982).
  %%CITATION = PRLTA,49,970;%%

\bibitem{pdecay1}
S. Weinberg, Phys. Rev. {\bf D26}, 287 (1982); N. Sakai and T.
Yanagida, Nucl. Phys. {\bf B197}, 533 (1982); S. Dimopoulos, S.
Raby and F. Wilczek, Phys. Lett. {\bf B112}, 133 (1982); J. Ellis,
D.V. Nanopoulos and S. Rudaz, Nucl. Phys. {\bf B202},
 43 (1982).

\bibitem{pdecay2}
P.~Nath, A.~H.~Chamseddine and R.~Arnowitt,
%``Nucleon Decay In Supergravity Unified Theories,''
Phys.\ Rev.\ D {\bf 32}, 2348 (1985);
%%CITATION = PHRVA,D32,2348;%%
 %``Nucleon Decay Branching Ratios In Supergravity SU(5) Guts,''
  Phys.\ Lett.\  B {\bf 156}, 215 (1985);
  %%CITATION = PHLTA,B156,215;%%
 J. Hisano, H. Murayama and T. Yanagida, Nucl. Phys. {\bf B402}, 46 (1993);
 T. Goto, T. Nihei and J. Arafune, Phys. Rev. {\bf D52}, 505
(1995); K.~S.~Babu and S.~M.~Barr,
%``Proton decay and realistic models of quark and lepton masses,''
Phys.\ Lett.\ B {\bf 381}, 137 (1996);
%[arXiv:hep-ph/9506261].
%%CITATION = HEP-PH 9506261;%%
 P.~Nath and R.~Arnowitt,
  %``LIMITS ON PHOTINO AND SQUARK MASSES FROM PROTON LIFETIME IN SUPERGRAVITY
  %AND SUPERSTRING MODELS,''
  Phys.\ Rev.\  D {\bf 38}, 1479 (1988);
  %%CITATION = PHRVA,D38,1479;%%
T. Goto and T. Nihei, Phys. Rev. {\bf D59}, 115009 (1999).

\bibitem{mp}
  H.~Murayama and A.~Pierce,
  %``Not even decoupling can save minimal supersymmetric SU(5),''
  Phys.\ Rev.\  D {\bf 65}, 055009 (2002)
  [arXiv:hep-ph/0108104].
  %%CITATION = PHRVA,D65,055009;%%

\bibitem{dmr}
  R.~Dermisek, A.~Mafi and S.~Raby,
  %``SUSY GUTs under siege: Proton decay,''
  Phys.\ Rev.\  D {\bf 63}, 035001 (2001)
  [arXiv:hep-ph/0007213].
  %%CITATION = PHRVA,D63,035001;%%

\bibitem{Kobayashi:2005pe}
  K.~Kobayashi {\it et al.}  [Super-Kamiokande Collaboration],
  %``Search for nucleon decay via modes favored by supersymmetric grand
  %unification models in Super-Kamiokande-I,''
  Phys.\ Rev.\  D {\bf 72}, 052007 (2005)
  [arXiv:hep-ex/0502026].
  %%CITATION = PHRVA,D72,052007;%%

\bibitem{Yao:2006px}
  W.~M.~Yao {\it et al.}  [Particle Data Group],
  %``Review of particle physics,''
  J.\ Phys.\ G {\bf 33}, 1 (2006).
  %%CITATION = JPHGB,G33,1;%%

\bibitem{bps}
B.~Bajc, P.~Fileviez Perez and G.~Senjanovic,
%``Proton decay in minimal supersymmetric SU(5),''
Phys.\ Rev.\ D {\bf 66}, 075005 (2002);
%[arXiv:hep-ph/0204311].
%\bibitem{ew}
  D.~Emmanuel-Costa and S.~Wiesenfeldt,
  %``Proton decay in a consistent supersymmetric SU(5) GUT model,''
  Nucl.\ Phys.\  B {\bf 661}, 62 (2003)
  [arXiv:hep-ph/0302272].
  %%CITATION = NUPHA,B661,62;%%

\bibitem{textures}
  P.~Nath,
  %``Textured Minimal and Extended Supergravity Unification and Implications for
  %Proton Stability,''
  Phys.\ Lett.\  B {\bf 381}, 147 (1996)
  [arXiv:hep-ph/9602337];
  %%CITATION = PHLTA,B381,147;%%
  %``Hierarchies and Textures in Supergravity Unification,''
  Phys.\ Rev.\ Lett.\  {\bf 76}, 2218 (1996)
  [arXiv:hep-ph/9512415].
  %%CITATION = PRLTA,76,2218;%%

\bibitem{cp}
T.~Ibrahim and P.~Nath,
%``Effects of large CP phases on the proton lifetime in supersymmetric
%unification,''
Phys.\ Rev.\ D {\bf 62}, 095001 (2000).
%[arXiv:hep-ph/0004098]
%%CITATION = HEP-PH 0004098;%%

\bibitem{suppress}
K.~S.~Babu and S.~M.~Barr,
%``Natural suppression of Higgsino mediated proton decay in supersymmetric
%SO(10),''
Phys.\ Rev.\ D {\bf 48}, 5354 (1993);
%[arXiv:hep-ph/9306242]
%%CITATION = HEP-PH 9306242;%%
Z. Chacko and R.N. Mohapatra, Phys.Rev. {\bf D59}, 011702 (1999);
Phys. Rev. Lett. {\bf 82}, 2836 (1999); Z. Berezhiani, Z.
Tavartkiladze and M. Vysotsky, hep-ph/9809301; I.~Gogoladze and
A.~Kobakhidze,
%``Natural suppression of d = 5 operator induced proton decay in  supersymmetric grand unified theories,''
Phys.\ Atom.\ Nucl.\  {\bf 60} (1997) 126 [Yad.\ Fiz.\  {\bf 60N1}
(1997) 136];
%[arXiv:hep-ph/9610389].
 T.~Dasgupta, P.~Mamales and P.~Nath,
  %``Effects of gravitational smearing on predictions of supergravity grand
  %unification,''
  Phys.\ Rev.\  D {\bf 52}, 5366 (1995)
  [arXiv:hep-ph/9501325];
  %%CITATION = PHRVA,D52,5366;%%
G. Altarelli, F. Feruglio, and I. Masina, JHEP {\bf 0011}, 040
(2000); Q. Shafi and Z. Tavartkiladze, Phys. Lett. {\bf B487}, 145
(2000); N. Maekawa, Prog. Theor. Phys. {\bf 106}, 401 (2001);
K.~Turzynski,
%``On suppressing the Higgsino-mediated proton decay in SUSY SO(10) GUT's,''
JHEP {\bf 0210} (2002) 044.
%[arXiv:hep-ph/0110282].

\bibitem{Dutta:2004zh}
  B.~Dutta, Y.~Mimura and R.~N.~Mohapatra,
  %``Suppressing proton decay in the minimal SO(10) model,''
  Phys.\ Rev.\ Lett.\  {\bf 94}, 091804 (2005)
  [arXiv:hep-ph/0412105].
  %%CITATION = PRLTA,94,091804;%%

\bibitem{Ibrahim:1998je}
  T.~Ibrahim and P.~Nath,
  %``The neutron and the lepton EDMs in MSSM, large CP violating phases, and
  %the cancellation mechanism,''
  Phys.\ Rev.\  D {\bf 58}, 111301 (1998);
%  T.~Ibrahim and P.~Nath,
  %``Large CP phases and the cancellation mechanism in EDMs in SUSY, string  and
  %brane models,''
  Phys.\ Rev.\  D {\bf 61}, 093004 (2000)
  [arXiv:hep-ph/9910553].
  %%CITATION = PHRVA,D61,093004;%%
 For a review see,   T.~Ibrahim and P.~Nath,
  ``CP violation from standard model to strings,''
  arXiv:0705.2008 [hep-ph].
  %%CITATION = ARXIV:0705.2008;%%

\bibitem{Arnowitt:1993pd}
  R.~Arnowitt and P.~Nath,
  %``Testing supergravity grand unification at future accelerator and
  %underground experiments,''
  Phys.\ Rev.\  D {\bf 49}, 1479 (1994)
  [arXiv:hep-ph/9309252].
  %%CITATION = PHRVA,D49,1479;%%

  \bibitem{45plet}
 G.~Segre and H.~A.~Weldon,
  %``SU(5) Theories With Both Proton Stability And Cosmological Baryon Number
  %Generation,''
  Phys.\ Rev.\ Lett.\  {\bf 44}, 1737 (1980);
  %%CITATION = PRLTA,44,1737;%%
 H.~S.~Tsao,
  %``M (W) = M (Z) Cos Theta-W In SU(5),''
  Phys.\ Rev.\  D {\bf 24}, 791 (1981);
  %%CITATION = PHRVA,D24,791;%%
 L.~Arnellos and W.~J.~Marciano,
  %``Hydrogen - Anti-Hydrogen Oscillations, Double Proton Decay And Grand
  %Unified Theories,''
  Phys.\ Rev.\ Lett.\  {\bf 48}, 1708 (1982);
  %%CITATION = PRLTA,48,1708;%%
P.~Eckert, J.~M.~Gerard, H.~Ruegg and T.~Schucker,
  %``Minimization Of The SU(5) Invariant Scalar Potential For The
  %Fortyfive-Dimensional Representation,''
  Phys.\ Lett.\  B {\bf 125}, 385 (1983);
  %%CITATION = PHLTA,B125,385;%%
    I.~Dorsner and P.~F.~Perez,
  %``Unification versus proton decay in SU(5),''
  Phys.\ Lett.\  B {\bf 642}, 248 (2006)
  [arXiv:hep-ph/0606062];
  %%CITATION = PHLTA,B642,248;%%
  P.~F.~Perez,
  %``Supersymmetric Adjoint SU(5),''
  arXiv:0705.3589 [hep-ph].
  %%CITATION = ARXIV:0705.3589;%%

\bibitem{Haba:2006qp}
  N.~Haba and T.~Ota,
  %``Vanishing dimension five proton decay operators in the SU(5) SUSY GUT,''
  arXiv:hep-ph/0608244.
  %%CITATION = HEP-PH/0608244;%

\bibitem{pdecay3}
V. Lucas and S. Raby,  Phys. Rev. {\bf D54}, 2261 (1996); Phys.
Rev. {\bf D55}, 6986 (1997).

\bibitem{pdecay4}
  K.S. Babu, J.C. Pati and F. Wilczek, Nucl. Phys. {\bf B566} 33 (2000);
 %``Suggested new modes in supersymmetric proton decay,''
Phys.\ Lett.\ B {\bf 423}, 337 (1998).
%[arXiv:hep-ph/9712307].
%%CITATION = HEP-PH 971230

\bibitem{bgns1}
  K.~S.~Babu, I.~Gogoladze, P.~Nath and {R.~M.~Syed},
 % ``A unified framework for symmetry breaking in $SO(10)$,''
  Phys.\ Rev.\ D {\bf 72}, 095011 (2005)
  [arXiv: hep-ph/0506312].
  %%CITATION = HEP-PH 0506312;%%

\bibitem{bgns2}
  K.~S.~Babu, I.~Gogoladze, P.~Nath and {R.~M.~Syed},
  %``Fermion mass generation in SO(10) with a unified Higgs sector,''
  Phys.\ Rev.\  D {\bf 74}, 075004 (2006)
  [arXiv:hep-ph/0607244].
  %%CITATION = PHRVA,D74,075004;%%

\bibitem{ms}
R.N. Mohapatra and B. Sakita, Phys. Rev. {\bf D21}, 1062 (1980).

\bibitem{wilczek}
F. Wilczek and A. Zee, Phys. Rev. {\bf D25}, 553 (1982).

\bibitem{ns}
 P.~Nath and {R.~M.~Syed},
 %``Analysis of couplings with large tensor representations in SO(2N) and  proton decay,''
Phys.\ Lett.\ B {\bf 506}, 68 (2001);
% [arXiv: hep-ph/0103165];
%%CITATION = HEP-PH 0103165;%%
%``Complete cubic and quartic couplings of 16 and 16-bar in SO(10) unification,''
Nucl.\ Phys.\ B {\bf 618}, 138 (2001);
% [arXiv: hep-th/0109116]; ``Coupling the supersymmetric 210 vector multiplet to matter in SO(10),''
Nucl.\ Phys.\ B {\bf 676}, 64 (2004);
% [arXiv: hep-th/0310178];
%%CITATION = HEP-TH 0310178;%%
 R. M. Syed,
 %in {\it Themes in Unification: Pran Nath Festschrift}, ed. by G. Alverson and M.T. Vaughn, (World Scientific, Singapore).
 arXiv: hep-ph/0411054;
 %%CITATION = HEP-PH 0411054;%%
  %``Couplings in SO(10) Grand Unification,''  Ph.D. Thesis, Northeastern University.
   arXiv: hep-ph/0508153.
  %%CITATION = HEP-PH 0508153;%%

\bibitem{ns1}
  P.~Nath and {R.~M.~Syed},
  %``Couplings of vector-spinor representation for $SO(10)$ model building,''
  JHEP {\bf 0602}, 022 (2006).
%  [arXiv: hep-ph/0511172];
  %%CITATION = HEP-PH 0511172;%%

\bibitem{Arkani-Hamed:2004fb}
  N.~Arkani-Hamed and S.~Dimopoulos,
  %``Supersymmetric unification without low energy supersymmetry and  signatures
  %for fine-tuning at the LHC,''
  JHEP {\bf 0506}, 073 (2005)
  [arXiv:hep-th/0405159].
  %%CITATION = JHEPA,0506,073;%%

\bibitem{Kors:2004hz}
  B.~Kors and P.~Nath,
  %``Hierarchically split supersymmetry with Fayet-Iliopoulos D-terms in  string
  %theory,''
  Nucl.\ Phys.\  B {\bf 711}, 112 (2005)
  [arXiv:hep-th/0411201].
  %%CITATION = NUPHA,B711,112;%%

\bibitem{p-future}
  K.~Nakamura,
 % ``Hyper-Kamiokande: A next generation water Cherenkov detector,''
  Int.\ J.\ Mod.\ Phys.\ A {\bf 18} (2003) 4053;
  %%CITATION = IMPAE,A18,4053;%%
  C.~K.~Jung,
 % ``Feasibility of a next generation underground water Cherenkov detector:
  UNO,''
  arXiv:hep-ex/0005046;
  M.~V.~Diwan {\it et al.},
  %``Megaton modular multi-purpose neutrino detector for a program of  physics
  %in the Homestake DUSEL,''
  arXiv:hep-ex/0306053;
  %%CITATION = HEP-EX 0306053;%%
A. de Bellefon et al,
%"MEMPHYS: A large scale water Cerenkov detector at Frejus",
contribution to the CERN strategy committee, Orsay 30/01/06;
  L.~Mosca,
%  ``A European megaton project at Frejus,''
  Nucl.\ Phys.\ Proc.\ Suppl.\  {\bf 138} (2005) 203;
  %%CITATION = NUPHZ,138,203;%%
  A.~Rubbia,
  %``Review of massive underground detectors,''
  arXiv:hep-ph/0407297;
  D.~B.~Cline,
  %``A unique detector for proton decay and neutrino oscillations study (LANNDD)
  %for a USA DUSEL,''
  arXiv:astro-ph/0506546;
  %%CITATION = ASTRO-PH 0506546;%%
L. Oberaurer,
%"low energy neutrino astrophysics", talk given at NNN05, Aussois, April 2005.
 http://nnn05.in2p3.fr; See also T. Marrodan Undagoitia \textit{et
al}, Phys. Rev. D{\bf 72}, 075014 (2005);
  A.~Bueno {\it et al.},
  %``Nucleon decay searches with large liquid argon TPC detectors at shallow
  %depths: Atmospheric neutrinos and cosmogenic backgrounds,''
  JHEP {\bf 0704}, 041 (2007)
  [arXiv:hep-ph/0701101].
  %%CITATION = JHEPA,0704,041;%%

\bibitem{Lebedev:2007hv}
  O.~Lebedev, H.~P.~Nilles, S.~Raby, S.~Ramos-Sanchez, M.~Ratz, P.~K.~S.~Vaudrevange and A.~Wingerter,
  %``The Heterotic Road to the MSSM with R parity,''
  arXiv:0708.2691 [hep-th].
  %%CITATION = ARXIV:0708.2691;%%

\end{thebibliography}
\end{document}